\documentclass[lettersize,journal]{IEEEtran}
\usepackage{amsmath,amsfonts}
\usepackage{algorithmic}
\usepackage{algorithm}
\usepackage{array}
\usepackage[caption=false,font=normalsize,labelfont=sf,textfont=sf]{subfig}
\usepackage{textcomp}
\usepackage{stfloats}
\usepackage{url}
\usepackage{verbatim}
\usepackage{graphicx}
\usepackage{cite}
\begin{document}

\title{Negative-Voltage-Enabled Energy Efficient Nonvolatile Memories And In-Memory Computing Based On 2D Piezoelectric Transistors}

\author{Jeffry Victor, Sumeet K. Gupta
\thanks{This work was supported by the Army Research Office and Laboratory of Physical Sciences under Grant W911NF-19-1-04

The authors are with the Elmore Family School of Electrical and Computer Engineering, Purdue University, West Lafayette, IN 47907 USA (e-mail: louis8@purdue.edu; guptask@purdue.edu).
}
}

\markboth{Journal of \LaTeX\ Class Files}%
{Shell \MakeLowercase{\textit{et al.}}: A Sample Article Using IEEEtran.cls for IEEßE Journals}


\maketitle

\begin{abstract}
Piezoelectric FET (PeFET) is a promising non-volatile-memory (NVM) device that integrates a piezoelectric (PE)/ferroelectric (FE) capacitor with a 2D transistor. It uses the polarization of the FE capacitor for bit-storage and strain-induced bandgap change of the 2D channel during read. Previous PeFET-based NVM designs have shown immense promise in achieving high density and energy-efficiency compared to SRAM. However, a key limitation of these designs is that they must trade-off integration density to enhance energy-efficiency or augment the memory functionality with in-memory computing (IMC). In this work, we show that the unique structure of the PeFET presents an appealing opportunity to counter these limitations, thereby simultaneously achieving high density, high energy-efficiency, and IMC-compatibility. First, we highlight the key reasons for the limited energy-efficiency of the previous PeFET designs. Based on these insights, we propose two flavors of PeFET memories that utilize negative voltage (NeVo) to reduce the major energy-consuming components significantly. Compared to 6T-SRAM (prior PeFET-based NVMs), the proposed designs achieve substantial reductions in energy, lowering read energy to 0.08×(0.03×) and write energy to 0.19×(0.53×), respectively. We then leverage these cells to implement IMC primitives, such as addition, subtraction, and multiply-and-accumulate (MAC), achieving 0.03× the energy consumption of prior PeFET-based designs.
\end{abstract}


\section{Introduction}
\IEEEPARstart{M}odern computing systems systems face two major bottlenecks: the slowdown of technology scaling and the growing memory wall problem \cite{technology_scaling, memory_wall}. On one hand, the slowdown in scaling has limited performance gains in static random access memory (SRAM) caches, limiting the overall system efficiency \cite{technology_scaling}. On the other hand, the memory wall—the high energy and latency of moving data between processors and memory—poses a serious challenge for data-intensive workloads such as deep neural networks (DNNs) \cite{memory_wall, tim_dnn}.

To overcome these limitations, emerging NVM technologies are being explored to design more efficient memory macros that not only perform standard memory operations but also support in-memory computing (IMC), where computation is carried out within the array to reduce data transfer overhead.

In the context of new technologies, NVMs such as resistive RAM (RRAM) \cite{geniex, magic}, magnetic RAM (MRAM) \cite{shubham_jain, jeffry_icecs}, and ferroelectric-based memories \cite{sandeep_ternary, aziz_fefets} offer advantages such as higher density (than SRAMs) and zero standby leakage but also face challenges. For instance, RRAM and MRAM suffer from energy-intensive, current-driven write mechanisms \cite{geniex}. While ferroelectric memories utilize electric field-driven write mechanisms to improve energy-efficiency, they have their own challenges, such as high write voltages (ferroelectric transistors or FeFETs), destructive read (ferroelectric RAMs or FERAMs), variability, and retention-related issues \cite{sandeep_ternary, aziz_fefets}. 

Piezoelectric FETs (PeFETs) mitigate these challenges while retaining the benefits of electric field-driven write \cite{niharika_ted}. PeFETs utilize a piezoelectric (PE)/ferroelectric (FE) capacitor (such as lead zirconate titanate, \textit{PZT-5H}) and use the FE polarization for bit storage. The FE/PE capacitor is coupled to a 2D transition metal dichalcogenide (TMD) transistor. The 2D TMD channel provides a strain-induced bandgap change mechanism, which is used during the read operation (discussed later).

 Prior works \cite{niharika_ted, niharika_fnano} have explored PeFETs in the context of memory design as well as IMC. The work in \cite{niharika_ted} introduced PeFET-based NVMs, achieving up to 5.8× reduction in read energy and 2.2× reduction in area compared to 6T-SRAM. The work in \cite{niharika_fnano} implemented Signed Ternary Precision (STP) Multiply-and-Accumulate (MAC) (used to accelerate Ternary DNNs) using PeFETs (utilizing the unique strain-based characteristics) to achieve over 6x energy savings compared to SRAM.

While the previously proposed PeFET NVMs \cite{niharika_ted, niharika_fnano} already achieve substantial energy savings over SRAM in both cache designs and IMC applications, a substantial portion of their energy consumption stems from charging highly capacitive bit-lines. Further, these designs typically trade off integration density for IMC operations. To address these limitations, we propose two novel PeFET-based NVM designs that utilize the unique strain-based read mechanism in conjunction with negative voltage (NeVo)-based biasing schemes to enhance the energy efficiency of PeFET NVMs further and incorporate IMC-compatibility without much area penalty. Besides the IMC of STP-MAC (with higher energy efficiency than the previous PeFET design), we propose IMC of addition and subtraction leveraging the strain-based read. 

The key contributions of this work are as follows:
\begin{itemize}
    \item We identify the limitations of existing PeFET NVMs and demonstrate how the unique properties of PeFETs can be leveraged to overcome these challenges.
    \item We propose two novel PeFET NVM designs that achieve significant energy savings by utilizing negative voltage biasing to eliminate the charging of read bit-lines (\textit{RBLs}) and write bit-lines (\textit{WBLs}).
    \item We utilize the proposed PeFET NVMs to support IMC primitives such as addition, subtraction, and STP-MAC.
    \item We benchmark the proposed PeFET NVMs against SRAM and existing PeFET-based NVMs for read, write, and IMC operations.
\end{itemize}

\section{Background}
\subsection{PeFET structure and operation}

\begin{figure}[!t]
\centering
\includegraphics[height = 2.25in, width=0.85\columnwidth]{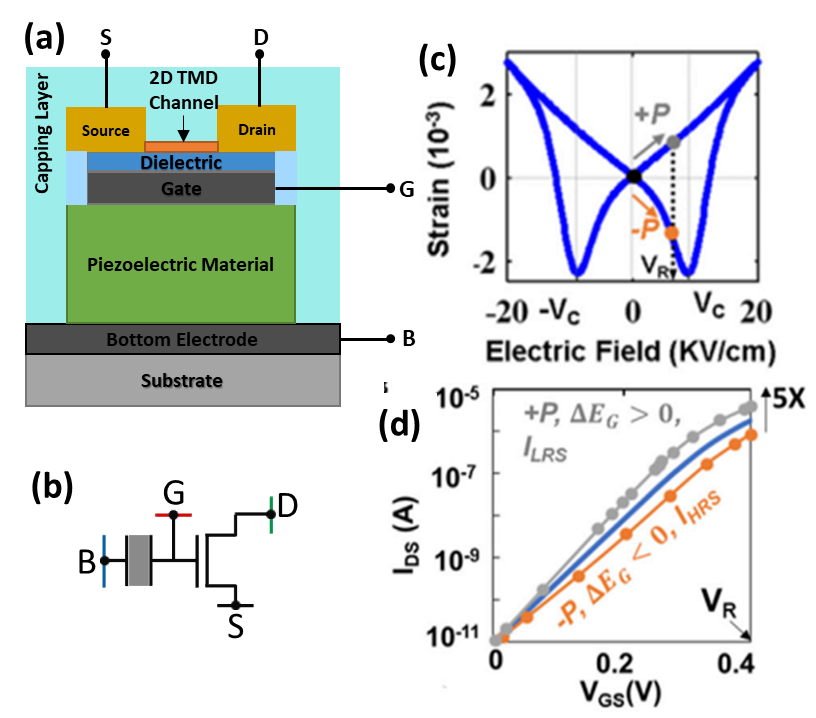}
\caption{(a) PeFET device (b) PeFET schematic (c) Strain vs \textit{$V_{GB}$} for +ve \textit{$V_{GB}$} (d) IV characteristics for +ve \textit{$V_{GB}$}}
\label{fig_device}
\end{figure}

PeFET is a four-terminal structure (Fig. \ref{fig_device}a\&b) comprising a FE/PE capacitor between the gate (\textit{G}) and back (\textit{B}) terminals and a 2D TMD transistor on top of the PE capacitor. The PE/FE capacitor consists of \textit{PZT-5H}. The channel is made of monolayer $MoS_2$. The PeFET leverages the polarization retention capability of the  (PE)/(FE) material for non-volatile bit storage, where the polarization states (\textit{+P}/\textit{-P}) represent stored bit values (1/0). 

For the write operation, gate-to-back voltage (\textit{$V_{GB}$}) is applied across PE, whose magnitude is greater than that of the coercive voltage ($V_{C}$). To set the polarization to \textit{+P}, we apply $V_{GB} > +V_C$, and to reset the polarization to  \textit{-P}, we apply $V_{GB} < -V_C$.

For reading, a positive $V_{GB}$ $<$ $V_C$ is applied. The PE material expands if it stores \textit{+P} and contracts if it stores \textit{-P}. This induces a positive (for \textit{+P}) or negative (for \textit{-P}) strain in the 2D TMD channel (Fig. \ref{fig_device}c), leading to a corresponding decrease (for \textit{+P}) or increase (for \textit{-P}) in its bandgap. As a result, the drain current ($I_{DS}$) corresponds to a low-resistance state (\textit{LRS}) for \textit{+P} and a high-resistance state (HRS) for \textit{-P}, as illustrated in Fig. \ref{fig_device}d \cite{niharika_ted}. Since the $V_{GB}$ remains below \textit{$V_C$}, this process enables a non-destructive read operation. 

\begin{table}[!t]
\caption{PeFET Parameters\label{tab:device}}
\centering
\begin{tabular}{|c|c|}
\hline
\textbf{Parameter} & \textbf{Value}\\
\hline
Gate Length & 20 nm\\
\hline
Channel Length & 30 nm\\
\hline
PE Length & 100 nm\\
\hline
PE Width & 150 nm\\
\hline
VDD & 0.8 V\\
\hline
Area of active $MoS_2$ & 30x20 $nm^2$\\
\hline
Thickness of $MoS_2$ & 0.65 nm\\
\hline
Mobility of monolayer $MoS_2$ & 90 $cm^2/Vs$\\
\hline
Coefficient of bandgap change in $MoS_2$ & 1.5 eV/GPa\\
\hline
Out-of-plane coupling coefficient of PZT-5H & 650 pm/V\\
\hline
\end{tabular}
\end{table}

To accurately capture device behavior, we model the polarization-electric-field response of the FE/PE capacitor using the Preisach model \cite{presiach}, calibrated with experimental data \cite{experimental}. The pressure transduced from the PE layer to the 2D-TMD is simulated using the 3D structure of the PeFET in the COMSOL Multiphysics suite. Furthermore, the pressure-induced bandgap change is self-consistently coupled with the Verilog-A-based 2D TMD FET model \cite{s2ds} that reflects the dynamically changing band gap on the charge/potential of the 2D TMD channel. We use a capacitive network-based model, similar to the Stanford 2D Semiconductor (S2DS) transistor model \cite{s2ds}, to represent a back-gated 2D-FET accurately. The key parameters of the PeFET device are summarized in Table \ref{tab:device}, and further details on the device modeling can be found in \cite{niharika_ted, niharika_fnano}.

\subsection{PeFET based NVMs}
\begin{figure}[!t]
\centering
\includegraphics[height=4in,width=0.8\columnwidth]{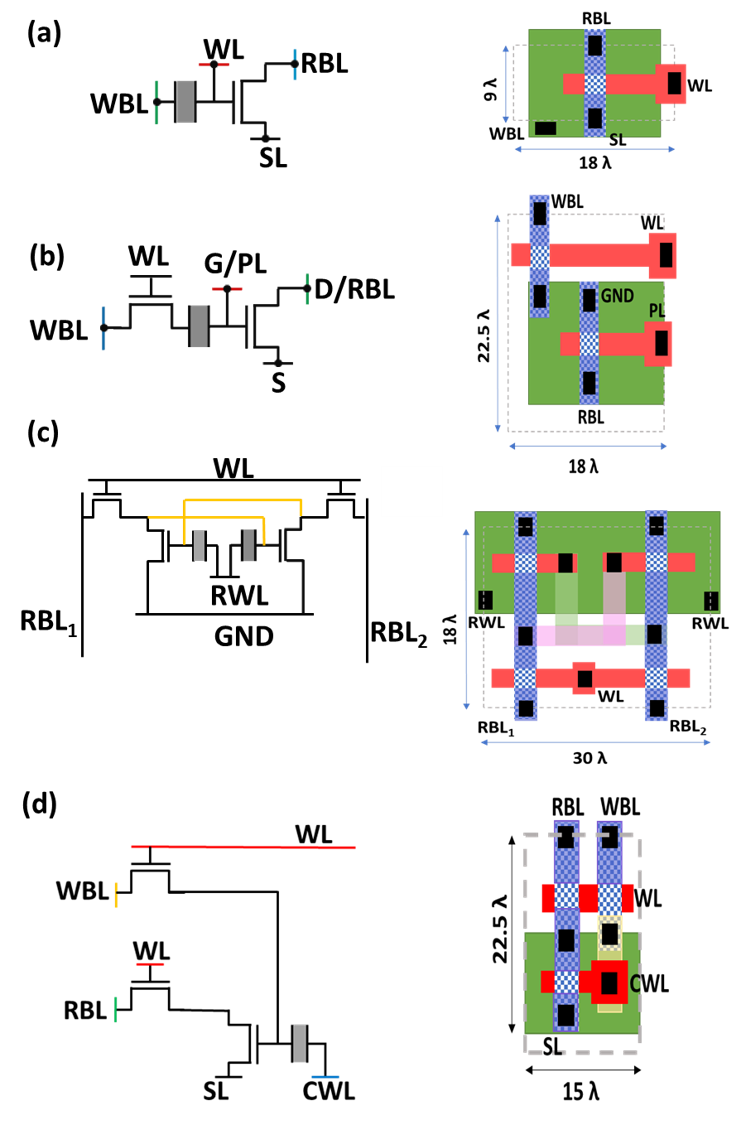}
\caption{Schematic and layouts of (a) HD (b) 1T-1P (c) CC and (d) 2T-1P NVMs.}
\label{fig_prev_nvms}
\end{figure}

The works in \cite{niharika_ted, niharika_fnano} have explored several PeFET-based NVMs, including High Density (HD) NVM, 1 Transistor–1 PeFET (1T-1P) NVM, Cross-Coupled (CC) NVM, and 2 Transistor–1 PeFET (2T-1P) NVM. HD NVM (shown in Fig. \ref{fig_prev_nvms}a) is a compact cell consisting of only 1 PeFET. Its compact layout leads to an efficient read operation but suffers from high write energy and latency arising from the charging of highly capacitive piezoelectrics in the half-accessed cells \cite{niharika_ted}.

The 1T-1P NVM (Fig. \ref{fig_prev_nvms}b) was proposed to enable more efficient write operations than the HD NVM by incorporating an access transistor between the PeFET and the write bit-line (\textit{WBL}), preventing the charging of PE capacitors in half-accessed cells \cite{niharika_ted}. While this design reduces write latency and energy compared to the HD NVM, it comes at the cost of increased cell area. The \textit{B} terminal of the PeFET in the 1T-1P configuration cannot be shared across a row, leading to a larger area footprint. This increased area, in turn, makes the read operation less efficient than the HD NVM \cite{niharika_ted}.

To mitigate the charging of PE capacitors in unaccessed cells with milder area penalty (compared to 1T-1P), the CC NVM was proposed \cite{niharika_ted}. Cross-coupling the two PeFETs (Fig. \ref{fig_prev_nvms}c) allows sharing of the \textit{B} contact along the row, conserving cell area. This leads to a comparable read efficiency and improved write efficiency with respect to the HD NVM \cite{niharika_ted}. However, the downside of this design is the reduced read distinguishability, which arises mainly as a result of the cross-coupling.

Compared to SRAM, PeFET NVMs offer a significantly smaller area footprint and lower read energy \cite{niharika_ted}. The latter is primarily due to their compact layout, which reduces bitline capacitance and associated charging energy. However, these benefits come at the expense of higher write energy and increased read and write latencies \cite{niharika_ted} (primarily due to the effect of PE capacitances).

Further, to extend the benefits of PeFETs to IMC, the work in \cite{niharika_fnano} introduced the 2T-1P cell (Fig. \ref{fig_prev_nvms}d). This design used two access transistors per bit-cell to enable the computation of STP-MAC operations used in Ternary DNNs. However, the enhanced functionality comes at the cost of an increase in the cell area and read energy (compared to the HD cell).

\begin{figure}[!t]
\centering
\includegraphics[width=3.25in, height=1.25in]{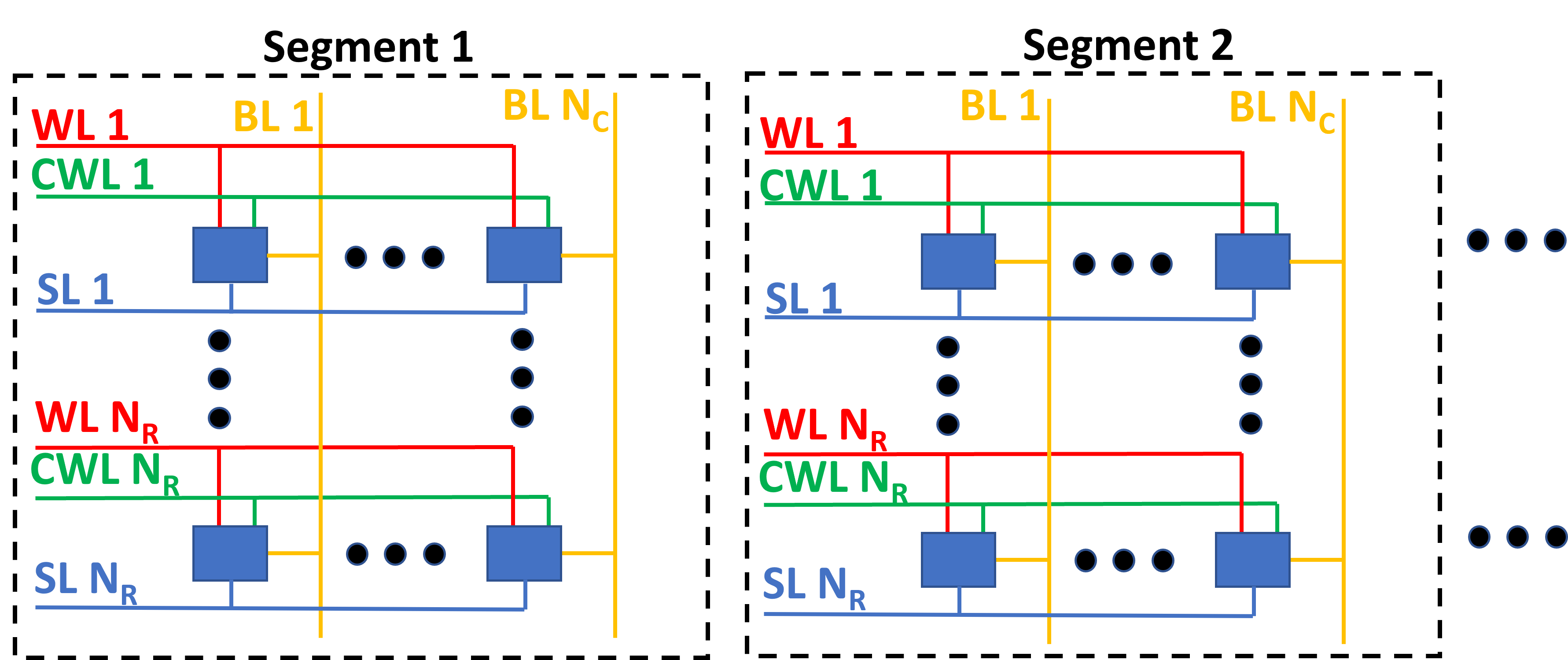}
\caption{Segmented Memory Array}
\label{fig_seg}
\end{figure}

Another key aspect of PeFET array design is the use of segmentation to reduce energy and latency consumed in charging BLs and PEs associated with unaccessed columns and rows, respectively \cite{niharika_ted}. Similar to segmentation in FeRAMs \cite{feram_seg}, the array is divided into column-based segments, each comprising $N_r$ rows and $N_c$ columns, and equipped with local word lines (\textit{WLs}), compute word lines (\textit{CWLs}), and source lines (\textit{SLs}) that activate only the cells within that segment (Fig.\ref{fig_seg}). This confines BL charging and discharging to only the $N_c$ BLs within the accessed segment, while leaving unaccessed segments undisturbed. Moreover, segmentation ensures that PE capacitance is charged only for the addressed bits within the activated segment, rather than all the bits in that row. This reduces both read and write latency. Although segmentation introduces some peripheral overhead, the energy and latency savings from avoiding unnecessary switching of unaccessed \textit{BLs} and PE capacitances makes it a highly effective design trade-off.

\subsection{Energy Analysis of PeFET NVMs}
\begin{figure}[!t]
\centering
\includegraphics[width=2.25in, height=1.25in]{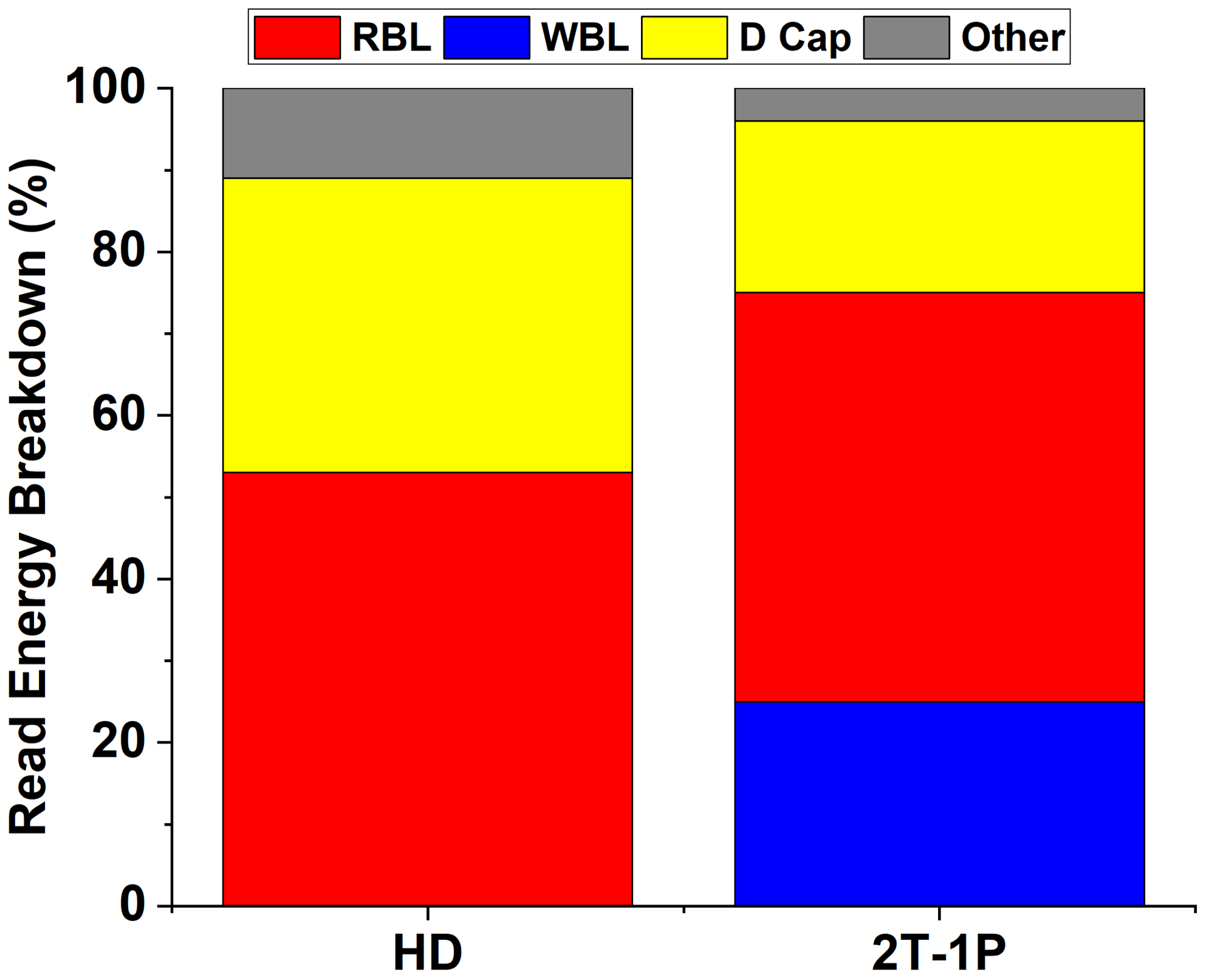}
\caption{Analyzing the contribution of different components in current-sensing read energy of HD and 2T-1P normalized to each NVM}
\label{fig_rbl}
\end{figure}

Although previous PeFET designs show promising results, the trade-offs between area, read-write energy, and IMC compatibility limit their benefits. With the motivation to understand these trade-offs further, we conduct a detailed analysis of the energy components to explain the bottlenecks. Our analysis of the current-sensing read operation (Fig. \ref{fig_rbl}) shows that a major portion of the energy consumption of PeFET bit-cells arises from charging the read /write bit-lines (\textit{RBLs}/\textit{WBLs}), and the drain terminals (i.e., the drain capacitance \textit{D Cap}) of the bit-cells (connected to these lines - see (Fig. \ref{fig_prev_nvms})) within a segment. These lines span the length of the entire memory array and have high capacitance, making their charging highly energy-intensive. As shown in Fig. \ref{fig_rbl}, these components account for 86\% of the read energy in the HD cell and 96\% in the 2T-1P cell \cite{niharika_ted, niharika_fnano}. 

In this work, we eliminate these energy components by leveraging the unique read mechanism of PeFETs through a NeVo-based biasing scheme combined with bit-cell redesign. Applying this approach to HD and 2T-1P NVMs results in NeVo HD and NeVo 2T-1P NVMs, respectively. Furthermore, the NeVo approach enables the NeVo HD NVM to support IMC primitives, which were not possible in the HD NVM. 

\subsection{Generation of negative voltage}
Since the proposed designs require negative voltages, we briefly discuss the methods used to generate negative voltages in memory sub-systems. Multiple prior works in the literature have explored the generation of a negative voltage to be used for memory arrays \cite{sram_negative_bl,sram_write_assist}. Mukhopadhyay et al utilize a capacitor connected to each bit line of the SRAM \cite{sram_negative_bl}. This capacitor is initially charged, followed by an application of 0V on its positive terminal. This results in a negative voltage on its other terminal, which, in turn, is applied to the bit line. Chang et al have used a capacitor in a similar way, but connected it to a driver that eventually applies the negative voltage onto the bit line \cite{sram_write_assist}. These works show that the generation of negative voltages can be achieved on-chip efficiently and at the high frequency required by caches.  

\section{NeVo HD PeFET NVM}
\subsection{NeVo HD PeFET Bit-cell and Array} 
\begin{figure}[!t]
\centering
\includegraphics[width=\columnwidth]{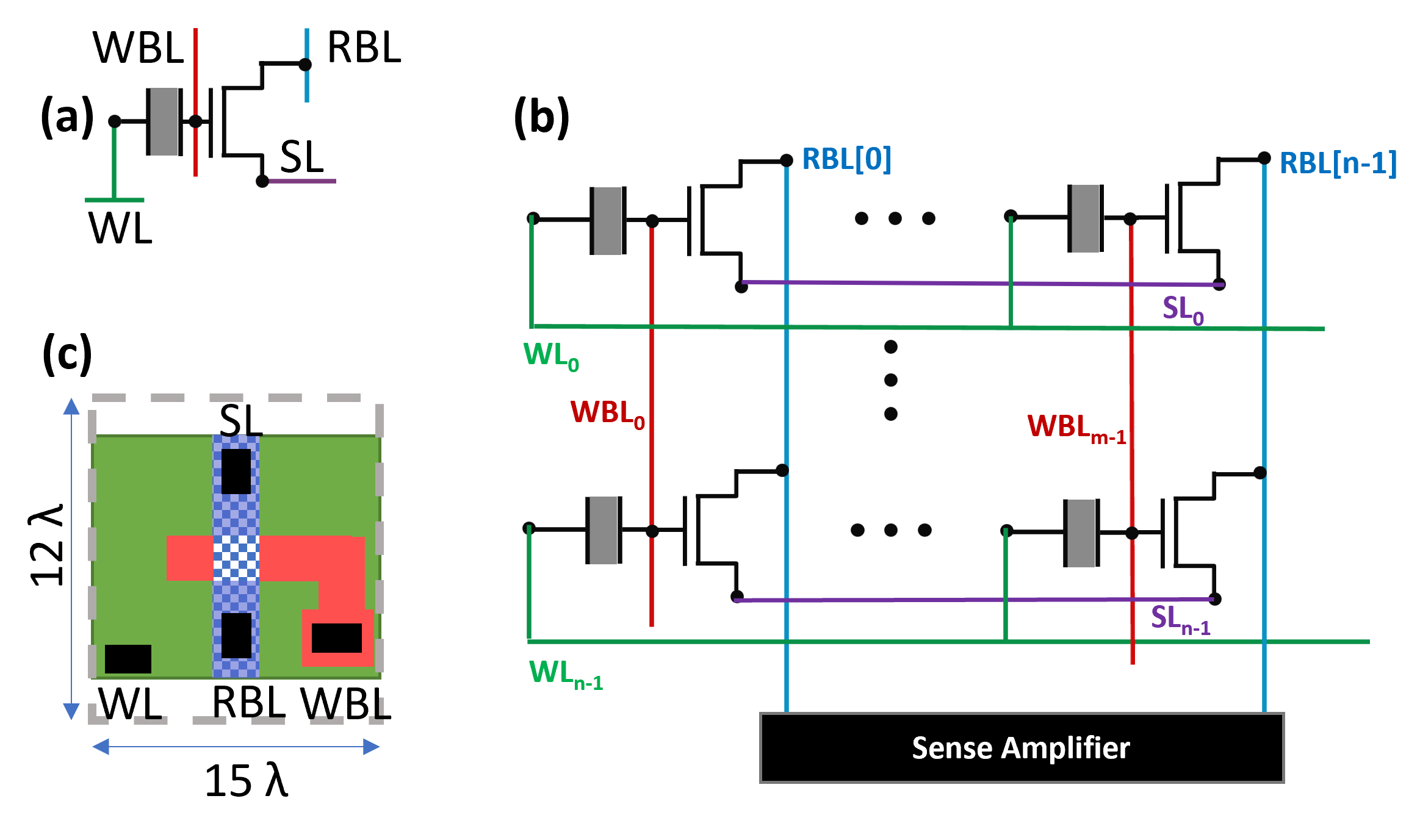}
\caption{NeVo HD (a) Schematic (b) Array and (c) Layout}
\label{fig_nevo_HD}
\end{figure}

Fig. \ref{fig_nevo_HD}a shows the proposed NeVo HD design, which features a single PeFET without any access transistor. The \textit{G} terminal of the PeFET is connected to the write bit-line (\textit{WBL}), routed vertically, while the source (\textit{S}) and drain (\textit{D}) terminals are connected to the horizontally routed source line (\textit{SL}) and the vertically routed \textit{RBL}, respectively. The \textit{B} contact is connected to the wordline \textit{WL} running along the row. \textit{WL} and \textit{WBL} establish a cross-point connection to facilitate write operations (Fig. \ref{fig_nevo_HD}b), as discussed later. 

Note, in the HD cell (Fig. \ref{fig_prev_nvms}a) proposed in \cite{niharika_ted}, the \textit{G} and \textit{S} terminals were connected horizontally to the \textit{WL} and \textit{SL}, while the \textit{D} and \textit{B} terminals were connected vertically to \textit{RBL} and \textit{WBL}, respectively. This design enables vertical sharing of the PE capacitor, thereby minimizing cell height. In contrast, the proposed NeVo HD (Fig.\ref{fig_nevo_HD}c)  allows sharing PE across cells in a row. In the vertical direction, additional spacing is required between adjacent PEs along the column, leading to a moderately increased cell height and bit-line capacitances compared to the HD design. Nevertheless, this trade-off enables NeVo HD in the elimination of high-energy-consuming components during read, as discussed next.

\subsection{Read Operation based on Current Sensing} 

Let us first consider current-based sensing and discuss  the read biasing in HD and NeVo HD designs. In the HD design \cite{niharika_ted}, read is initiated by  applying \textit{$V_{DD}/2$} to \textit{WL} and grounding  \textit{WBL} and \textit{SL}. The \textit{RBL} is raised to {$V_{DD}$}, and the read current is sensed at \textit{RBL} \cite{niharika_ted}. In NeVo HD, we ground \textit{WBL} and \textit{RBL} while applying {$-V_{DD}/2$} to the \textit{WL} and \textit{SL}. This approach achieves the necessary bias conditions for read (that is, \textit{$V_{GB} = V_{DD}/2$}, \textit{$V_{GS}=V_{DD}/2$}, and \textit{$V_{DS}=V_{DD}/2$}) without raising the voltage of \textit{RBL}. 

Recall, the motivation to design NeVo PeFET bit-cells and arrays is to save the bit-line charging energy, which comprises a significant component in the previously proposed PeFET designs (Fig. \ref{fig_rbl}). By grounding the vertically connected \textit{RBL}, NeVo HD avoids the cost of charging RBL \textit{of each accessed bit}. Note that to establish the necessary read biasing conditions without charging \textit{RBL}, NeVo HD charges the horizontally connected \textit{WL} and \textit{SL}. This energy expenditure is comparatively minimal and is amortized amongst the bits in a word/row because these lines are \textit{shared by all accessed bits}. 

To further illustrate the energy savings achieved by averting the charging of \textit{RBL}, consider the difference in charged capacitance between the NeVo HD and HD arrays. Suppose,  \textit{$N_c$} bits are read from a segment with \textit{$N_c$} columns and \textit{$N_r$} rows. While the HD design activates all \textit{$N_c$} \textit{RBLs} during a read, the NeVo HD design activates only a single \textit{WL} and a single \textit{SL}. As a result, the NeVo HD design ends up charging significantly lower capacitance, leading to large energy savings.

It is noteworthy that the HD cell ( Fig. \ref{fig_prev_nvms}a) cannot utilize the NeVo approach to avoid RBL charging as it would entail grounding WL and applying negative voltage on \textit{WBL} to create the bias for read. Thus, while \textit{RBL} charging would be averted, it would be replaced by charging another vertically routed line, \textit{WBL}.  Also, note that to support segmentation within the arrays, both the \textit{SL} and \textit{WL} are segmented in NeVo HD, whereas only the \textit{WL} is segmented in HD. Segmenting these lines prevents the unintended charging of unaccessed bit-cells, thereby avoiding unnecessary BL charging. 

\begin{figure}[!t]
\centering
\includegraphics[width=\columnwidth]{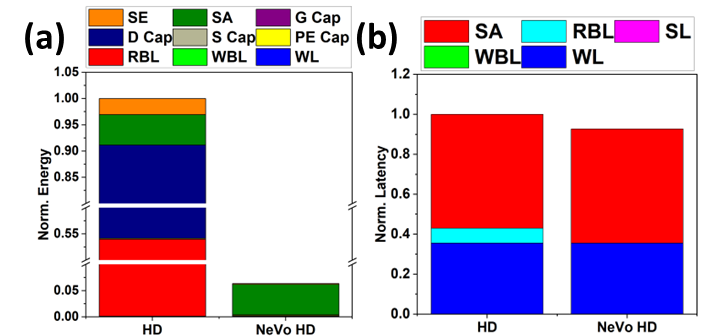}
\caption{Characterizing the Current-based Read Operation of NeVo HD and HD (a) Energy and (b) Latency Comparison.}
\label{fig_isense_HD}
\end{figure} 

In HD NVM, nearly 86\% of the energy is consumed in charging the drain and \textit{RBL} parasitic capacitances (as discussed in the previous section). By eliminating this overhead, NeVo HD reduces its energy to only 0.06× that of HD (Fig. \ref{fig_isense_HD}a), while maintaining comparable read latency (Fig. \ref{fig_isense_HD}b).

\subsection{Read Operation based on Voltage Sensing}

Although targeted mainly for current-sensing, let us now briefly discuss the implications of the NeVo approach for voltage sensing to bring out its limitations. Voltage sensing requires \textit{RBL} to be floated long enough to develop the target sense margin (100 mV) before sensing. In HD NVM, \textit{RBL} is first precharged to {$V_{DD}/2$} and then floated until the voltage differential is developed. In contrast, NeVo HD is designed to mitigate \textit{RBL} precharging; however, if \textit{RBL} was initially grounded and then floated, it would develop a negative voltage as \textit{SL} of the accessed row is biased at {$-V_{DD}/2$}. This negative \textit{RBL} voltage (negative $V_{GS}$) can increase the leakage in the unaccessed cells along the column, which, if large, could result in loss of read robustness. While this could be mitigated by utilizing smaller array sizes, preventing negative $V_{GS}$ altogether  requires \textit{RBL} in NeVo HD to be precharged to {$V_{DD}/2$} before floating. However, this removes the benefit of NeVo biasing—which eliminates the RBL precharge overhead—causing NeVo HD to incur about 1.14× the energy of HD NVM (at comparable latency) due to its taller layout.. Thus, the NeVo approach is more compatible with current-based sensing. Voltage-sensing either leads to increased leakage in the unaccessed cells (which needs careful optimization of read robustness) or is not able to capitalize on the basic advantage of the NeVo approach.  

Although NeVo HD NVM incurs a higher energy for voltage sensing than HD NVM considering standard memory read, it does have important benefits for IMC. We will discuss these in detail in Section VI.

\subsection{Write Operation in NeVo HD}

\begin{figure}[!t]
\centering
\includegraphics[height=0.15\textheight,width=0.85\columnwidth]{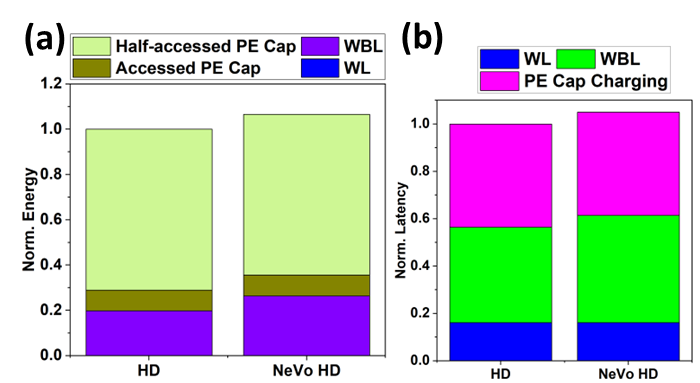}
\caption{Characterizing the Write operation of NeVo HD and HD (a)Energy and (b) Latency Comparison.}
\label{fig_wr_HD}
\end{figure}

 In NeVo HD, writing is performed by applying a two-phase signal ({$V_{DD}/2$} to {$-V_{DD}/2$}) on the \textit{WL}, while the \textit{WBL} is held at {$V_{DD}/2$} (or {$-V_{DD}/2$}) to write a \textit{+P} (or \textit{-P}) state. The HD design follows a similar approach, applying two-phase signals on the \textit{WL}, with the \textit{WBL} set to {$V_{DD}/2$} (or {$-V_{DD}/2$}) to write \textit{+P} (\textit{-P}) \cite{niharika_ted}. As shown in Fig.\ref{fig_wr_HD}a, the NeVo HD array consumes 1.08× the write energy of the HD array. This slight increase is attributed to higher energy consumption on the \textit{WBL}, due to the taller layout of the NeVo HD cell. Note that, since both designs are access-transistor-free, a significant portion of the write energy is expended in charging the half-accessed PE capacitances.

In terms of write latency (Fig.\ref{fig_wr_HD}b), both cells demonstrate comparable performance, with the NeVo HD array exhibiting 1.05× the latency of the HD array. This slight increase is primarily due to the longer time required to drive the \textit{WBL} in NeVo HD, a consequence of its taller layout.

In summary, NeVo HD significantly reduces current-sensing read energy by avoiding RBL charging. However, due to its taller layout, it incurs higher energy during voltage sensing compared to HD. In terms of write energy, both cells are comparable. Additionally, NeVo HD and HD exhibit similar latency across both read and write operations.

\section{NeVo 2T-1P PeFET NVM}
In this section, we extend the NeVo biasing strategy to the 2T-1P bit-cell through the proposed NeVo 2T-1P design, and evaluate its energy reduction potential. While the circuit schematic of NeVo 2T-1P NVM remains identical to the 2T-1P NVM \cite{niharika_fnano}, the key distinction lies in the applied biasing scheme.

\subsection{Read based on Current Sensing}

For read based on current sensing in the 2T-1P design,  \textit{WL} and \textit{RBL} are raised to \textit{$V_{DD}$} while  \textit{WBL} is raised to \textit{$V_{DD}/2$} \cite{niharika_fnano}. This leads to the required \textit{$V_{GB}$}, \textit{$V_{GS}$} and \textit{$V_{DS}$} biasing for read. However, this approach involves charging vertically connected lines (\textit{RBL}, \textit{WBL}), which, as discussed in the previous section, is energy-intensive. In contrast, the NeVo 2T-1P cell achieves the same biasing conditions more efficiently by grounding these vertical lines and instead biasing the horizontally routed \textit{CWL}, \textit{SL} and \textit{WL} to \textit{$-V_{DD}/2$}, \textit{$-V_{DD}/2$} and \textit{$V_{DD}$}, respectively.

\begin{figure}[!t]
\centering
\includegraphics[height=0.15\textheight, width=\columnwidth]{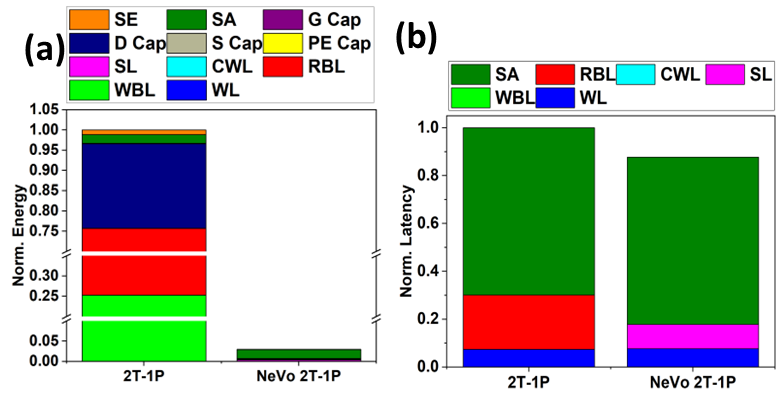}
\caption{Characterizing the Current sensing Read operation of NeVo 2T-1P and 2T-1P (a) Energy and (b) Latency Comparison.}
\label{fig_i_2t1p}
\end{figure}

NeVo 2T-1P achieves 0.03× the read energy of 2T-1P (Fig. \ref{fig_i_2t1p}a) by eliminating \textit{RBL}, \textit{WBL}, and drain capacitance charging, which accounts for 96\% of 2T-1P energy. It also reduces read latency to 0.87× by driving shorter segmented \textit{SL} and \textit{CWL} instead of long array-spanning \textit{RBL} and \textit{WBL}. The tall-and-thin layout of the 2T-1P further amplifies this advantage (Fig. \ref{fig_i_2t1p}b).

\subsection{Read based on Voltage Sensing in NeVo 2T-1P}
Let us now analyze the read energy consumption of NeVo 2T-1P compared to 2T-1P considering voltage-sensing. In voltage sensing, we pre-charge the \textit{RBL} in both 2T-1P and NeVo 2T-1P to \textit{$V_{DD}/2$}, then allow it to float for 0.5 ns i.e. the time needed for \textit{RBL} voltage to drop to establish sufficient sense margin (100 mV). This voltage drop is then sensed using a voltage sense amplifier. 

\begin{figure}[!t]
\centering
\includegraphics[width=\columnwidth]{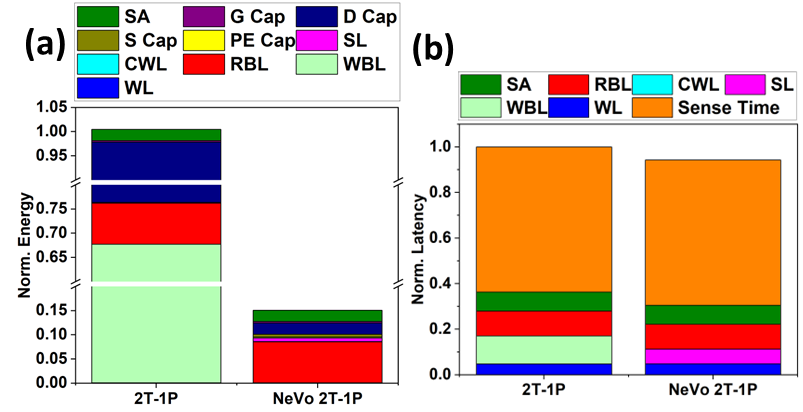}
\caption{Characterizing the Voltage sensing Read operation of NeVo 2T-1P and 2T-1P (a) Energy and (b) Latency Comparison.}
\label{fig_v_2t1p}
\end{figure}

NeVo 2T-1P consumes only 0.15× the read energy of 2T-1P (Fig. \ref{fig_v_2t1p}) by eliminating charging of \textit{WBL} and the associated drain capacitance. It achieves 0.94× the read latency of 2T-1P (Fig. \ref{fig_v_2t1p}b) due to the same reasons observed in the current-sensing read case. 

\subsection{Write Operation in NeVo 2T-1P}
In the 2T-1P design, \textit{WBL} is biased at \textit{$V_{DD}$} (\textit{GND}) in order to write \textit{+P} (\textit{-P}) \cite{niharika_fnano}. Then, the \textit{WL} is raised to \textit{$V_{DD}$} + 0.4V (i.e., 1.2 V). This boosted \textit{WL} voltage passes {$V_{DD}$} to the PeFET  across the access transistor. Finally,  \textit{CWL} is asserted with a two-phase signal (\textit{$V_{DD}$} in Phase 1 and \textit{GND} in Phase 2), which ensures that the magnitude of \textit{$V_{GB}$} is higher than that of the coercive voltage in either of the two phases depending on the data to be written. In contrast, in NeVo 2T-1P,  \textit{WBL} is raised to \textit{$V_{DD}/2$} (\textit{$-V_{DD}/2$}) in order to write \textit{+P} (\textit{-P}). Then, the \textit{WL} is raised to \textit{$V_{DD}$} + 0.4V. Finally, the \textit{CWL} is given a two-phase signal  (\textit{$V_{DD}/2$} in Phase 1 and \textit{$-V_{DD}/2$ in Phase 2}). 

\begin{figure}[!t]
\centering
\includegraphics[width=\columnwidth]{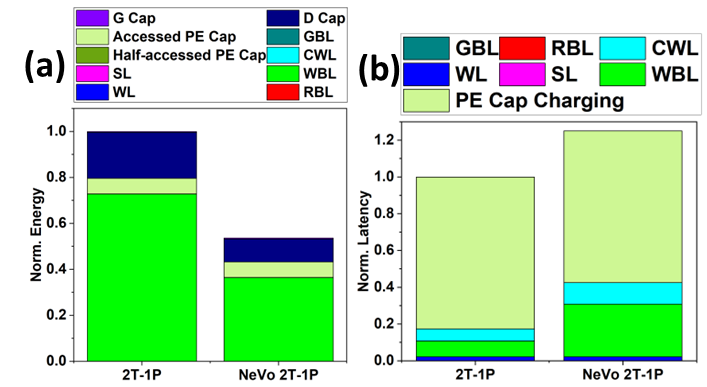}
\caption{Characterizing write operation of NeVo 2T-1P and 2T-1P (a) Energy and (b) Latency Comparison.}
\label{fig_wr_2t1p}
\end{figure}

From Fig. \ref{fig_wr_2t1p}a, we observe that the write energy in NeVo 2T-1P is 0.55x that of 2T-1P. The energy savings stem from the difference in WBL voltage levels: in 2T-1P, the \textit{WBLs} are driven to $V_{DD}$, whereas in NeVo 2T-1P, they are only driven to +/- $V_{DD}/2$. The higher voltage swing in 2T-1P leads to increased energy consumption associated with charging the \textit{WBLs}.

Let us now discuss the write latency comparisons. Fig. \ref{fig_wr_2t1p}b shows that most of the write latency is due to PE switching. The write latency of NeVo 2T-1P is 1.25x that of 2T-1P because \textit{WBL} and \textit{CWL} are driven to half the voltage in NeVo 2T-1P. The reduced \textit{$V_{GS}$} overdrive in the drivers of NeVo 2T-1P leads to its moderately higher latency compared to 2T-1P. 

To summarize, the NeVo 2T-1P NVM achieves significantly lower write and read energy compared to the conventional 2T-1P design—across both current-sensing and voltage-sensing modes. It also offers reduced read latency, while incurring a moderate increase in write latency.

\section{Benchmarking Memory Operations}
\begin{figure*}
\centering
\includegraphics[width=0.9\textwidth, height=1.85in]{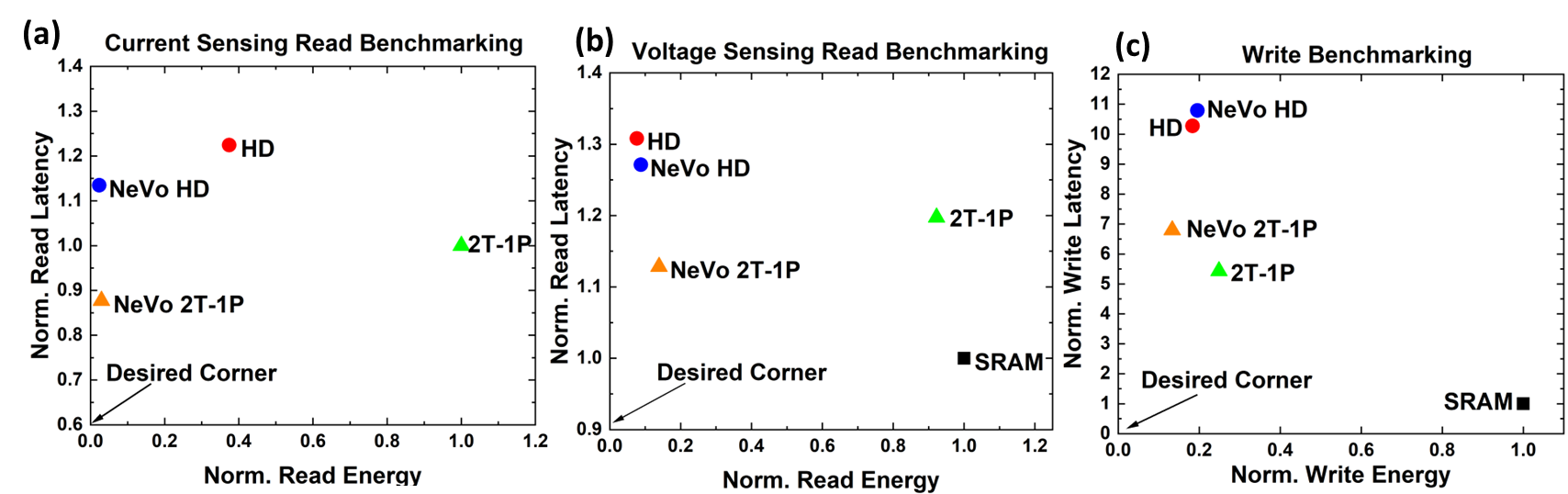}
\caption{Benchmarking NeVo HD and NeVo 2T-1P against previously proposed PeFET NVMs and 6T-SRAM in (a) Current-sensing read operation, (b) Voltage-sensing read operation, and (c) Write operation.}
\label{fig_mem_bmark}
\end{figure*}
In this section, we benchmark the memory operations of the proposed NeVo 2T-1P and NeVo HD designs against 6T-SRAM, HD, and 2T-1P. We utilize a 6T-SRAM with a silicon channel and planar CMOS transistors at 22 nm. We used Predictive Technology Models (PTM) at 22 nm \cite{PTM}. The SRAM is assumed to have a 30\% utilization and 70\% leakage, based on the results in \cite{niharika_ted}. Parameters such as area, and \textit{$V_{DD}$} of all technologies are reported in Table \ref{tab:circuit}. All designs are evaluated using a 512×512 array. The PeFET NVMs employ segmentation (32-column segments) to reduce energy and latency from charging unaccessed bitlines and PE capacitances. Since SRAM does not drive large PE capacitors, segmentation offers little benefit and adds unnecessary overhead. Hence, we compare against an unsegmented SRAM array, which is the typical design for SRAMs.

\begin{table}[!t]
\caption{Circuit-level parameters of SRAM and PeFET NVMs\label{tab:circuit}}
\centering
\resizebox{\columnwidth}{!}{%
\begin{tabular}{|c|c|c|c|c|c|c|c|}
\hline
\textbf{Metrics} & \textbf{Si 6T-SRAM} & \textbf{HD} & \textbf{2T-1P} & \textbf{NeVo HD} & \textbf{NeVo 2T-1P}\\
\hline
Array Size & 512x512 & 512x512 & 512x512 & 512x512 & 512x512\\
\hline
Bit-cell Area & 15.8 & 2.8 & 7.0 & 3.0 & 7.0\\
(MPxGP) & & & & &\\
\hline
VDD (V) & 0.8 & 0.8 & 0.8 & 0.8 & 0.8\\
\hline
\end{tabular}
}
{\small Metal Pitch = 64 nm. Gate Pitch = 90 nm.}
\end{table}

The area comparison in Table II shows that 2T-1P and NeVo 2T-1P occupy 0.44× the area of SRAM, due to lower number of transistors. NeVo 2T-1P and 2T-1P have identical areas since they share the same bit-cell schematic and differ only in their biasing. NeVo HD achieves further area savings, occupying 0.42× the area of 2T-1P due to the absence of access transistors. Overall, NeVo HD exhibits 0.19x the area of SRAMs. However, NeVo HD has 1.07× the area of HD, as discussed before.

\subsection{Read Comparison: Current-based Sensing}
Fig. \ref{fig_mem_bmark}a shows that NeVo HD and NeVo 2T-1P achieve much lower current-sensing energy than HD and 2T-1P by using negative-voltage biasing to avoid charging vertical WBLs and RBLs. HD and NeVo HD have higher read latency than 2T-1P and NeVo 2T-1P as they charge unaccessed PE capacitances due to the lack of access transistors. Note, here, SRAMs are not included since they typically utilize voltage-sensing. 

\subsection{Read Comparison: Voltage-based Sensing}   
Fig. \ref{fig_mem_bmark}b shows that all PeFET NVMs exhibit lower voltage-sensing energy than SRAM, albeit with slightly higher latency. The energy savings primarily stem from the significantly smaller cell area of PeFETs compared to SRAM. SRAM achieves lower latency through fast differential sensing, whereas PeFET NVMs use the slower single-ended sensing. Similar to current-sensing read, 2T-1P and NeVo 2T-1P achieve lesser latency than HD and NeVo HD in voltage-sensing read as well.

\subsection{Write Comparison}
Fig.\ref{fig_mem_bmark}c shows that PeFET NVMs exhibit lower write energy but higher latency compared to SRAM. The energy savings are mainly due to their smaller cell area, while the increased latency arises from the need to drive PE capacitances. Among the PeFET NVMs, NeVo 2T-1P exhibits lower write energy and latency compared to HD, NeVo HD and 2T-1P as it avoids charging unaccessed PE capacitances and uses lower voltage biases ($V_{DD}/2$). 2T-1P shows the highest write energy among all PeFET NVMs due to its large area and the need to charge vertically running \textit{WBLs} to $V_{DD}$.

Overall, among PeFET NVMs, NeVo 2T-1P offers high energy efficiency and lower latency for both read and write operations. However, NeVo HD and HD provides a significantly smaller area footprint compared to NeVo 2T-1P.

The significant read energy savings offered by NeVo HD and NeVo 2T-1P make them very attractive candidates for designing IMC primitives, which typically rely on efficient read operations. Thus, we explore the design of in-memory addition and subtraction in the next section and followed by the design of in-memory MAC using the proposed NeVo designs.

\section{Pefet Enabled In-Memory Addition \& Subtraction}
\begin{figure}[!t]
\centering
\includegraphics[width=\columnwidth]{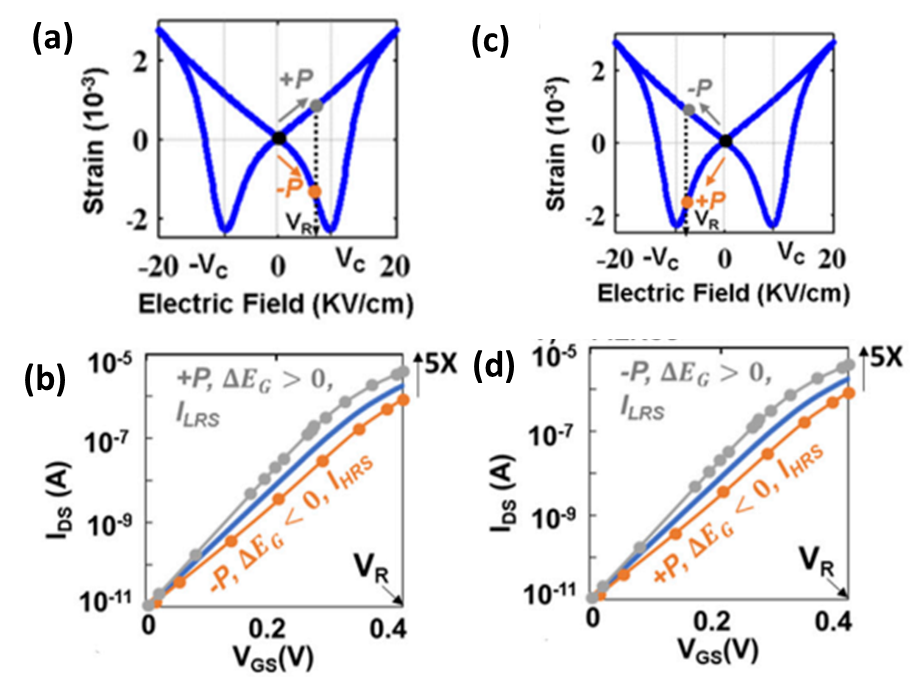}
\caption{(a) Strain vs \textit{$V_{GB}$} for +ve \textit{$V_{GB}$} (b) IV characteristics for +ve \textit{$V_{GB}$} (c) Strain vs \textit{$V_{GB}$} for -ve \textit{$V_{GB}$} (b) IV characteristics for -ve \textit{$V_{GB}$}}
\label{fig_device2}
\end{figure}
\subsection{Previous Works on IMC of Addition and Subtraction}
Several previous works have explored IMC for in-memory operands \cite{ambit, x_sram} based on multi-wordline assertion, but most of them are limited to commutative functions such as additions (and cannot handle non-commutative operation such as subtraction). There are a few exceptions. Works like \cite{supreet_sram} use a split-wordline 6T SRAM bitcell to
enable single-cycle two-bit read, which could further be used to implement both addition and subtraction functions, but this method is limited to differential cells with two access transistors. Another work, MAGIC \cite{magic} enables in-memory addition and subtraction in RRAM arrays but incurs a long latency due to the need for multiple intermediary cycles \cite{magic}. ADRA \cite{adra}, features single-cycle 2-bit read utilizing 2 wordline voltage levels and executes single-cycle addition and subtraction for a wide range of memory technologies, but at the cost of additional peripheral circuits such as three sense amplifiers (SAs) per sense line (\textit{SL}). In this work, we propose IMC designs using NeVo HD and NeVo 2T-1P, which execute both in-memory addition and subtraction, and address the limitations of the previous techniques.

\subsection{\textit{$V_{GB}$}-Polarity and Polarization-controlled current in PeFETs}
In NVM technologies such as ferroelectric memories, MRAM, RRAM etc., the order parameter/configuration of the device alone (such as polarization, magnetization, presence/absence of conducting filaments) determines whether the device is in the low resistance state (\textit{LRS}) or the high resistance state (\textit{HRS}). However, in PeFETs, the polarization of the FE/PE layer and the polarity of \textit{$V_{GB}$} dictates whether the device is in \textit{LRS} or \textit{HRS} \cite{niharika_ted, niharika_fnano}. 

As discussed before (in read operations), when we apply a positive $V_{GB}$ (that is less than $V_C$), the PE material expands if it stores \textit{+P} and contracts if it stores \textit{-P}. This expansion or contraction induces a corresponding positive or negative strain in the PE (Fig.\ref{fig_device2}a), which is transferred to the 2D TMD layer, resulting in a decrease (for \textit{+P}) or increase (for \textit{-P}) in its bandgap. As a result, the drain current ($I_{DS}$) corresponds to \textit{LRS} for \textit{+P} and \textit{HRS} for \textit{-P}, as illustrated in Fig. \ref{fig_device2}b \cite{niharika_ted}. 

But one also has a choice to apply a negative \textit{ $V_{GB}$} (such that \textit{$|V_{GB}|$} $<$ \textit{$|V_C|$}). In this case, the PE material contracts if it stores \textit{+P} and expands if it stores \textit{-P} (Fig. \ref{fig_device2}c). This leads to the bandgap of 2D TMD to increase (for \textit{+P}) or decrease (for \textit{-P}) - the opposite of the positive $V_{GB}$ case. Consequently, the drain current ($I_{DS}$) corresponds to \textit{HRS} for \textit{+P} and \textit{LRS} for \textit{-P}, as illustrated in Fig. \ref{fig_device2}d \cite{niharika_ted}. 

The dependence of current on \textit{$V_{GB}$}-polarity in addition to the stored polarization has been leveraged in prior works such as \cite{niharika_fnano} to implement in-memory MAC primitives. In this work, we exploit this mechanism in conjunction with the energy savings offered by the NeVo approach to design highly energy-efficient in-memory addition and  subtraction primitives. We also analyze the implications of NeVo designs in the in-memory MAC operations in the next section. 

\subsection{PeFET Enabled In-Memory Addition AND Subtraction}

\begin{figure}[!t]
\centering
\includegraphics[height=0.4\textheight,width=\columnwidth]{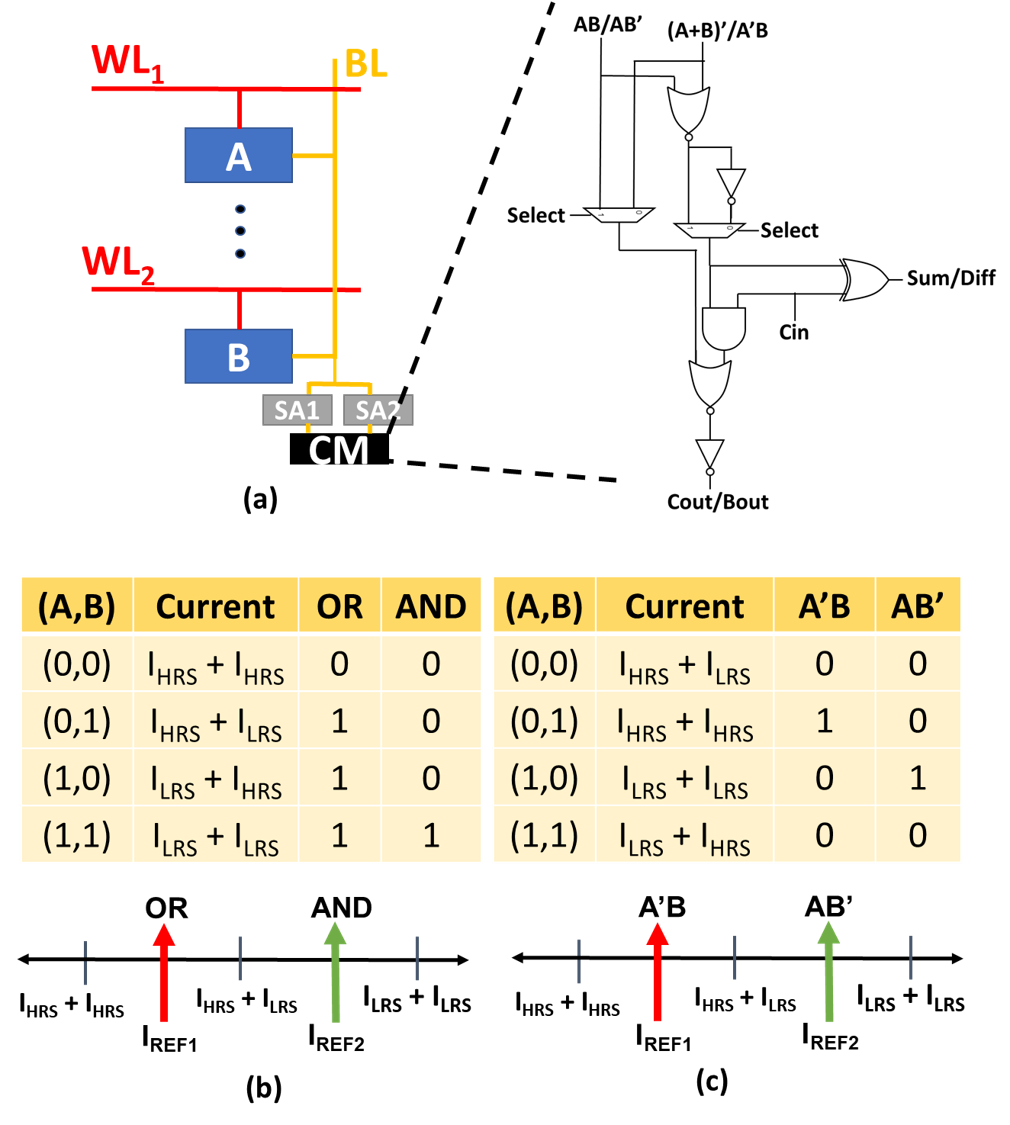}
\caption{(a) IMC capable memory array with 2 Sense Amplifiers (SAs) and Compute Module (CM). Mapping of input vectors, BL currents and outputs in in-memory (b) addition and (c) subtraction operations.}
\label{fig_sub_truth_table}
\end{figure}

Say, in two rows, two operands, \textit{A} and \textit{B} are stored (Fig. \ref{fig_sub_truth_table}a). If PeFETs in both rows are asserted with +ve \textit{$V_{GB}$} ($<V_c$), the currents generated in the \textit{RBL} follows the pattern shown in Fig. \ref{fig_sub_truth_table}b. The input vectors (0,0) and (1,1) correspond to 2\textit{$I_{HRS}$} and 2\textit{$I_{LRS}$}, respectively, while the (0,1) and (1,0) input vectors produce \textit{$I_{LRS}$}+\textit{$I_{HRS}$}. Using two sense amplifiers with appropriate reference currents (as shown in Fig. \ref{fig_sub_truth_table}b), we obtain bitwise OR (A+B), bitwise AND (AB), and their complements. By utilizing the compute module in the peripheral circuitry (as shown in Fig. \ref{fig_sub_truth_table}b), addition can be performed from these sense amplifier outputs. This approach—activating two word lines and processing the sense amplifier outputs using a compute module—is widely adopted in prior IMC works using SRAM, MRAM, RRAM, and other devices \cite{shubham_jain,x_sram}. We adopt the same strategy for in-memory addition using PeFETs. 

For in-memory subtraction (A-B), PeFETs in the row storing \textit{A} are asserted with +ve \textit{$V_{GB}$} while PeFETs in the other row storing \textit{B} are asserted with -ve \textit{$V_{GB}$} (such that $|V_{GB}<V_c|$). The resulting current follows the pattern shown in Fig. \ref{fig_sub_truth_table}c. The input vectors (0,1) and (1,0) correspond to 2\textit{$I_{HRS}$} and 2\textit{$I_{LRS}$} respectively, while the (0,0) and (1,1) input vectors produce \textit{$I_{LRS}$}+\textit{$I_{HRS}$} current. Using the same current references used for addition, we obtain bitwise A’B, bitwise AB’ and their complements, as shown in Fig. \ref{fig_sub_truth_table}c. By using these primitives and the compute module, in-memory subtraction can be performed. Thus, by exploiting the dependence of current on \textit{$V_{GB}$}-polarity and stored polarization, we are able to implement in-memory subtraction besides addition using PeFETs.

The discussion above highlights that the ability to support the same polarity of \textit{$V_{GB}$} across rows is required for in-memory addition, while the ability to support opposing polarities of \textit{$V_{GB}$} on different rows is crucial for enabling the in-memory subtraction operation. Among the PeFET memories, NeVo HD, NeVo 2T-1P and 2T-1P can support both positive and negative \textit{$V_{GB}$} on different rows. In the HD NVM \cite{niharika_ted}, the \textit{B} electrode is connected along the column (Fig. \ref{fig_prev_nvms}a). This implies that to enable opposite polarity \textit{$V_{GB}$} across two rows, one \textit{WL} must be set to $V_{DD}/2$ and the other to $-V_{DD}/2$. However, applying a negative voltage on the \textit{WL}, applies a negative potential to the gate of the PeFETs, preventing them from turning on. Thus, HD NVM can only support in-memory addition operation.

In contrast, NeVo HD and NeVo 2T-1P allow grounding of the \textit{G} terminal by driving the \textit{WBL} to 0 and using negative voltage on the \textit{S} terminal to turn the PeFET ON. The \textit{B} terminal, connected horizontally via the \textit{WL} in NeVo HD and \textit{CWL} in NeVo 2T-1P, can be driven to $+V_{DD}/2$ or $-V_{DD}/2$. Since the \textit{WL} and \textit{CWL} run horizontally, this enables applying opposite polarities of \textit{$V_{GB}$} across different rows, allowing the proposed NeVo designs to perform in-memory addition and subtraction. Although the 2T-1P NVM \cite{niharika_fnano} can also support opposite \textit{$V_{GB}$} polarity across rows, achieving the required \textit{$V_{GB}$}, \textit{$V_{GS}$}, and \textit{$V_{DS}$} biases requires asserting the vertically running \textit{RBL} and \textit{WBL}, making its in-memory addition and subtraction operations energy-inefficient (similar to the previous discussions on memory operations).

Next, we present the energy and latency results for the PeFET NVMs considering the in-memory subtraction operation. Note that energy and latency metrics remain identical for both addition and subtraction since these operations differ only in the application of ±$V_{DD}$/2 to the line connected to the \textit{B} terminal.

We benchmark the in-memory and near-memory subtraction operation across the three cells: NeVo HD, NeVo 2T-1P and 2T-1P, that support it. We also compare these results with the near-memory computing (NMC) of subtraction of all PeFET cells (including HD). The NMC process involves two read cycles, with the read data being written into registers after each cycle. This is  followed by reading the stored data from the registers and using a standard subtractor to perform the subtraction operation. 

\begin{figure}[!t]
\centering
\includegraphics[height=0.4\textheight,width=0.8\columnwidth]{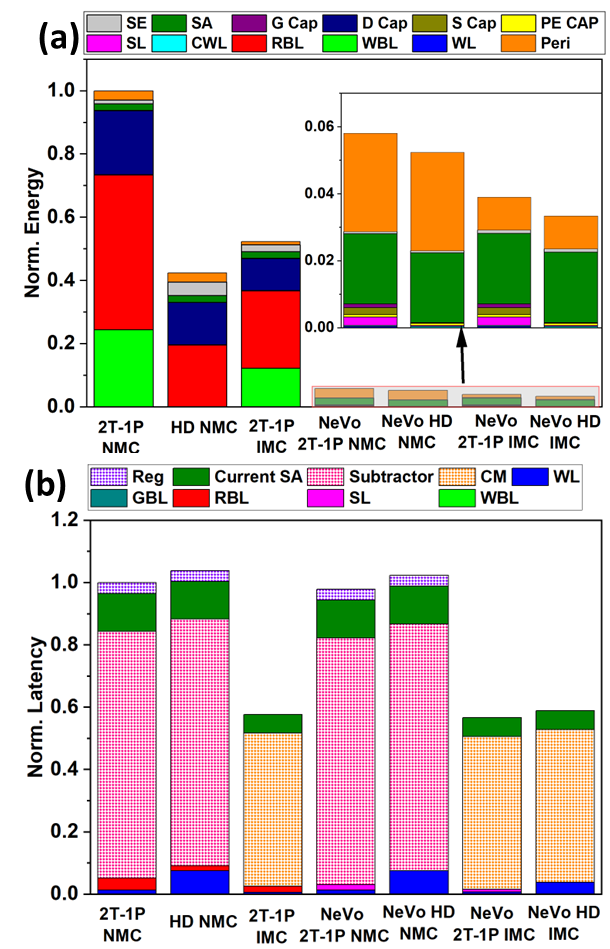}
\caption{Characterizing current-sensing subtraction (a) Energy and (b) Latency Comparison}
\label{fig_ISub}
\end{figure}

\subsection{Energy and Latency Comparison for Current-Sensing based Subtraction}
\textit{HD versus NeVo HD:} The energy comparison is shown in Fig.\ref{fig_ISub}a. NeVo HD NMC consumes 0.12× the energy of HD NMC.These significant savings stem from not asserting the vertical \textit{RBLs}—similar to the savings observed in read operations. Further, NeVo HD IMC consumes 0.63× the energy of NeVo HD NMC due to the simpler peripheral circuitry required for in-memory operations.

\textit{2T-1P versus NeVo 2T-1P:} As shown in Fig.\ref{fig_ISub}a, NeVo 2T-1P NMC consumes 0.06× the energy of 2T-1P NMC. These drastic savings for NeVo 2T-1P result from avoiding the charging of vertically connected lines-\textit{RBL} and \textit{WBL}. Further, NeVo 2T-1P IMC consumes 0.67× the energy of NeVo 2T-1P NMC, primarily due to simpler peripheral circuitry for IMC operations.

\textit{NeVo HD IMC versus NeVo 2T-1P IMC:} NeVo HD IMC consumes 0.85× the energy of NeVo 2T-1P IMC. This is because NeVo 2T-1P requires the assertion of larger number of lines (\textit{WL}, \textit{CWL}, and \textit{SL}) compared to NeVo HD (\textit{WL} and \textit{SL}).

In Fig. \ref{fig_ISub}b, the latencies of IMC and NMC operations across all cells are compared, showing that NMC operations take approximately 1.75× that of IMC. This is because IMC completes in a single cycle with simpler peripheral circuitry, while NMC spans two cycles and involves more complex peripheral circuitry. In both cases, the peripheral circuitry is a significant component of the latency, primarily due to the ripple-carry design of the subtractor (in NMC) and compute module (in IMC), which incurs significant delay from carry propagation across all columns in the accessed segment.

\begin{figure}[!t]
\centering
\includegraphics[height=0.4\textheight,width=0.8\columnwidth]{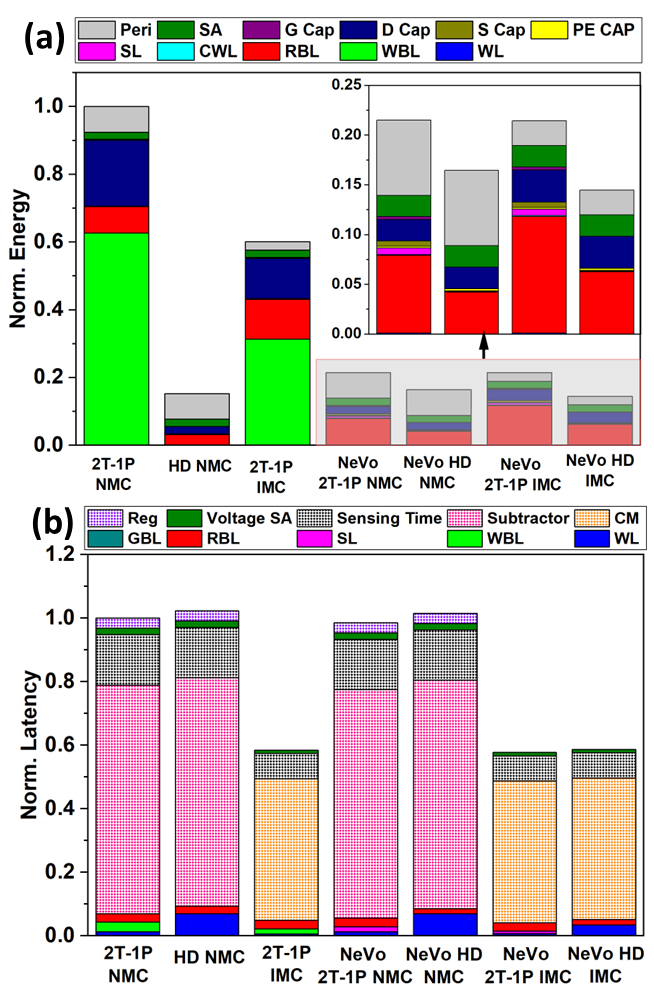}
\caption{Characterizing voltage-sensing subtraction (a) Energy and (b) Latency Comparison}
\label{fig_VSub}
\end{figure}

\subsection{Energy and Latency Comparison for voltage-sensing based Subtraction}

In this benchmarking, all PeFET NVMs have their \textit{RBLs} precharged to {$V_{DD}/2$}. The initial energy required for this precharge is excluded, as including it would disproportionately penalize near-memory operations that require two precharges. To provide a conservative estimate of in-memory energy savings, we consider only the energy needed to recharge the \textit{RBL} voltage drop that occurs during IMC and NMC operations, consistent with prior IMC studies \cite{adra}.

\textit{HD versus NeVo HD:} NeVo HD NMC consumes 1.05× the energy of HD NMC. This increase is primarily due to the taller layout of NeVo HD compared to HD, which results in higher RBL recharging energy. On the other hand, NeVo HD IMC consumes 0.87× the energy of NeVo HD NMC, with the savings attributed to simpler peripheral circuitry of IMC operations.

\textit{2T-1P versus NeVo 2T-1P:} NeVo 2T-1P NMC consumes 0.21× the subtraction energy of 2T-1P NMC. This substantial energy reduction arises from avoiding the assertion of WBL in NeVo 2T-1P. Additionally, NeVo 2T-1P IMC consumes approximately 0.99× the energy of NeVo 2T-1P NMC. While the peripheral circuitry in IMC consumes lesser energy than that of NMC, the overall IMC energy remains similar because IMC requires asserting two rows, which causes a greater voltage drop on RBL and results in higher recharging energy than NMC.

\textit{NeVo HD vs. NeVo 2T-1P:} NeVo HD IMC consumes 0.67× the energy of NeVo 2T-1P IMC. The primary reason for this energy reduction is the smaller layout height of NeVo HD, which leads to lower RBL charging energy.

From Fig. \ref{fig_VSub}b, voltage-sensing IMC and NMC subtraction show a similar latency trend as in current-sensing: NMC shows ~1.75× the latency of IMC due to single-cycle execution and simpler peripheral circuitry of IMC.

\subsection{Comparison with SRAMs}
\begin{figure}[!t]
\centering
\includegraphics[height=0.25\textheight, width=0.8\columnwidth]{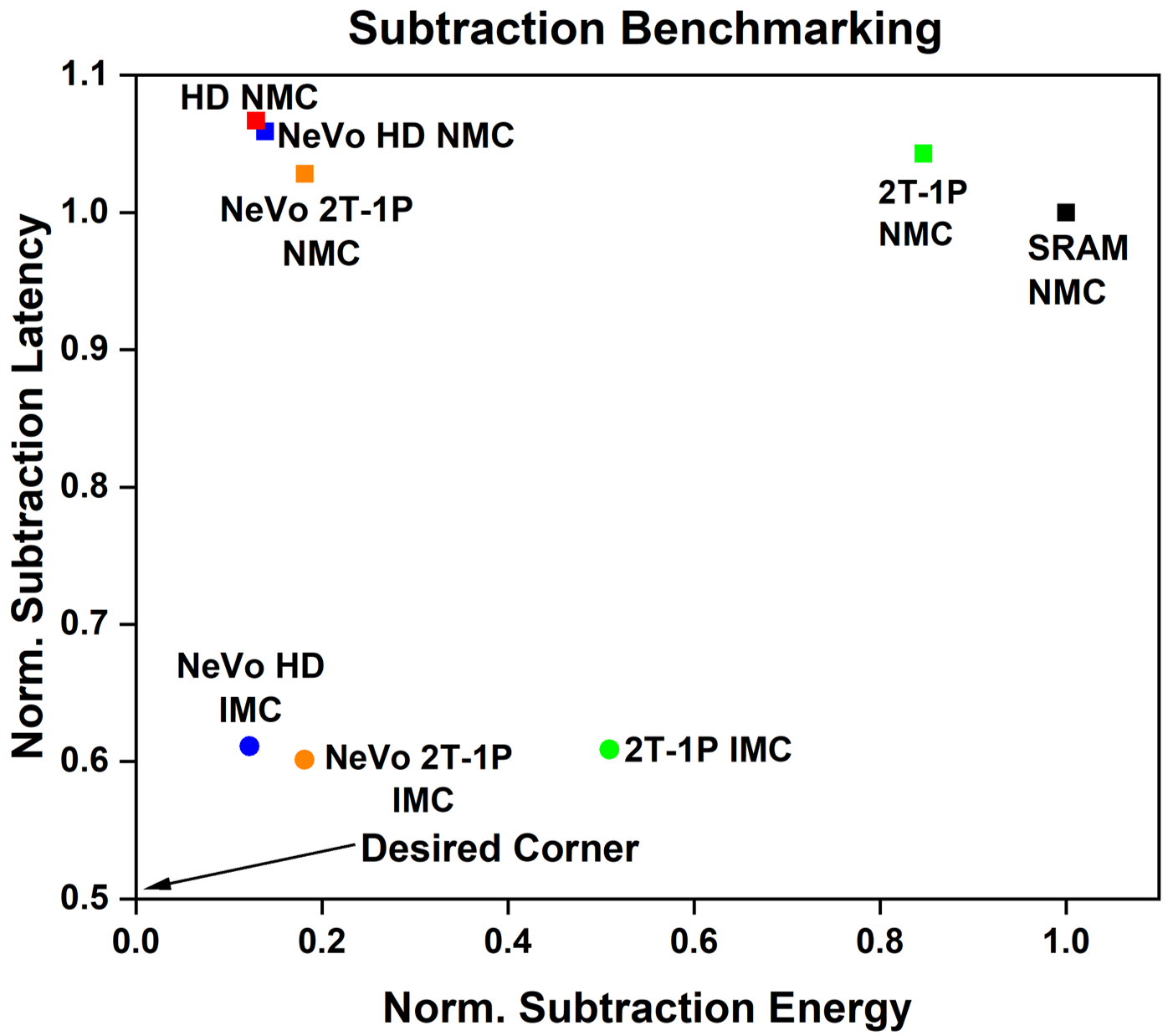}
\caption{Benchmarking PeFET NVMs and SRAM in voltage-sensing subtraction}
\label{fig_sub_comparison}
\end{figure}

Now, let us benchmark the proposed PeFET-based IMC and NMC with 6T SRAM-based NMC. This comparison is restricted to voltage-sensing, as implementing current-sensing NMC in 6T SRAMs is challenging (due to stability concerns \cite{x_sram}). From the results in Fig. \ref{fig_sub_comparison}, we observe that SRAM NMC has about the same latency as all the other NMC approaches but at a higher energy consumption. Since SRAM has the largest area, its NMC energy is higher compared to low-area cells such as NeVo 2T-1P, NeVo HD and HD. 2T-1P NMC energy consumption is close to that of SRAM NMC because of the high cost of charging the vertically running \textit{RBLs} and \textit{WBLs}. From our results in  Fig. \ref{fig_sub_comparison}, we also observe that all the IMC approaches are about half the latency of NMC approaches for the reasons discussed before.

\section{PeFET Enabled Ternary In-Memory MAC}
With the rise of DNNs, modern computing systems are increasingly constrained by the memory wall, struggling to meet the growing memory demands. Precision reduction, along with IMC, has become a key strategy for mitigating the memory wall in DNN accelerators \cite{precision1,precision2}. While 8-bit precision is standard in inference hardware, recent advancements suggest scaling down to binary precision \cite{precision1}, though at the cost of accuracy loss. Ternary precision networks, however, strike a balance, offering substantial energy savings with minimal accuracy degradation compared to higher-precision DNNs and significantly improved accuracy over binary DNNs \cite{tnn}. These advantages have driven growing interest in their hardware implementations \cite{edram, tim_dnn, sandeep_ternary}.
\begin{figure}[!t]
\centering
\includegraphics[height=0.4\textheight,width=0.8\columnwidth]{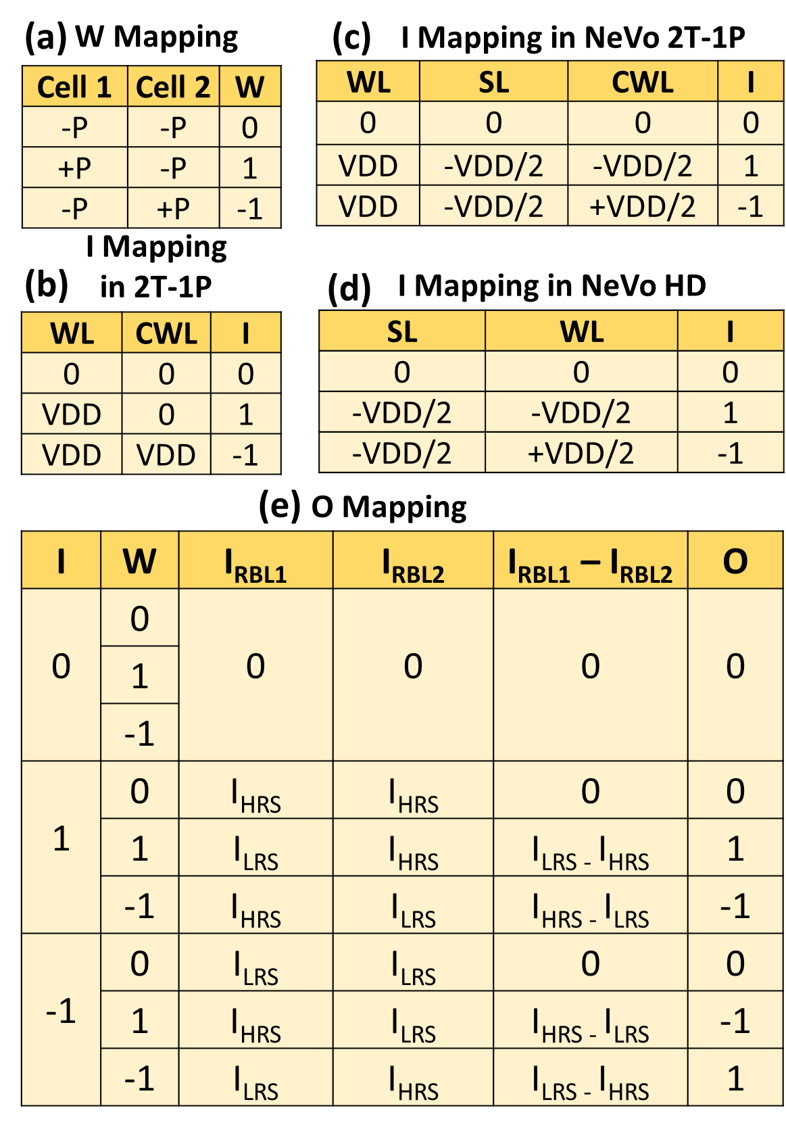}
\caption{Input (I), Weight (W) and Output (O) mapping for STP-MAC in PeFETs. (a) Weight mapping across PeFET NVMs. Input mapping in (b) 2T-1P (c) NeVo 2T-1P and (d) NeVo HD. (e) Out mapping across PeFET NVMs}
\label{fig_stp_truth_table}
\end{figure}
\subsection{Previous Works on In-Memory  Ternary MAC}
The work in \cite{edram} proposed eDRAM-based ternary CiM, but frequent refresh operations increase the energy burden on edge devices. Emerging technologies like RRAM and STT/SOT-MRAM offer high density and low leakage power for ternary networks but suffer from power-hungry current-driven writes, making them less suitable for energy-constrained environments \cite{niharika_fnano, sandeep_ternary}.

Most prior works use signed ternary weights with binary inputs, missing the accuracy benefits of pure signed ternary networks ({-1, 0, 1} for both weights and inputs). Recent studies have explored hardware accelerators for pure signed ternary networks using SRAM and ferroelectric transistor-based DNN architectures, demonstrating high parallelism and low energy consumption. However, SRAM-based ternary CiM faces area and leakage concerns \cite{tim_dnn}. Ferroelectric-based designs \cite{sandeep_ternary} reduce area and leakage but inherit other limitations of ferroelectric NVMs, as discussed earlier. 

To address these challenges, prior work \cite{niharika_fnano} proposed a PeFET-based STP-MAC using the 2T-1P cell. Leveraging the NeVo approach, which eliminates \textit{WBL}, \textit{RBL}, and drain-capacitance charging, we design an STP-MAC using the NeVo 2T-1P NVM to minimize these energy losses. Furthermore, we investigate the more compact NeVo HD cell for STP-MAC, anticipating that its reduced layout footprint may yield additional energy savings relative to NeVo 2T-1P. Note that STP-MAC cannot be implemented in HD, as it requires supporting opposite polarities of \textit{$V_{GB}$} across different rows (discussed below) —a capability that HD NVM lacks but NeVo HD provides.

\subsection{PeFET Enabled In-Memory STP-MAC}

STP-MAC in DNNs involves the multiplication of inputs and weights, which can take the values of {-1,0,1}. The weights are stored in the memory array in the form of polarization of the PeFETs. Each weight needs two bits to represent -1,0 and 1. So, we follow the weight mapping shown in Fig. \ref{fig_stp_truth_table}a. The weights {-1, 0, 1} are mapped to two adjacent cells (Cell 1 and Cell 2). If the PeFETs in both cells are initialized to \textit{-P}, then the weight mapped is weight '0'. If cell 1 is at \textit{+P} while cell 2 is at \textit{-P}, then it represents weight '1'. If it is vice-versa, then it stores weight '-1'. The NeVo HD, NeVo 2T-1P, and 2T-1P cells follow the same weight mapping.

The input mapping varies for each PeFET NVM. For the 2T-1P proposed in the work \cite{niharika_fnano}, the inputs are applied using the \textit{WL} and \textit{CWL} as shown in Fig. \ref{fig_stp_truth_table}b. The \textit{RBL} and \textit{WBL} are biased at \textit{$V_{DD}$} and \textit{$V_{DD}/2$} respectively. To apply input '0', both \textit{WL} and \textit{CWL} are grounded. To apply +1 (-1), the \textit{WL} is raised to \textit{$V_{DD}$} and the \textit{CWL} is driven to \textit{GND} (\textit{$V_{DD}$}), so that the voltage difference between the \textit{WBL} and the \textit{CWL} leads to \textit{$V_{GB}$} of \textit{$V_{DD}/2$} (\textit{$-V_{DD}/2$}). 

For NeVo 2T-1P,  \textit{WBL} and \textit{RBL} are grounded to avert the energy costs associated with charging them (as discussed earlier). Hence, \textit{WL}, \textit{SL}, and \textit{CWL} are used to supply the inputs. For input=0, all three lines are grounded. To map inputs = '1' ('-1'),  \textit{WL} is biased at \textit{$V_{DD}$}, \textit{SL} at \textit{$-V_{DD}/2$}, and \textit{CWL} at \textit{$-V_{DD}/2$} (\textit{$V_{DD}/2$}). This generates \textit{$V_{GB}$} of \textit{$V_{DD}/2$} (\textit{$-V_{DD}/2$}). 

In the NeVo HD cell, the \textit{WBL} and \textit{RBL} are grounded while the \textit{SL} and \textit{WL} are used to apply input signals as shown in Fig. \ref{fig_stp_truth_table}d. When both the \textit{WL} and \textit{SL} are grounded, the input applied is '0'. If the \textit{SL} is at \textit{$-V_{DD}/2$} while the \textit{WL} is at \textit{$-V_{DD}/2$} (\textit{$V_{DD}/2$}), the input mapped is '1' (-1).  

The truth table for the STP-MAC operation is shown in Fig. \ref{fig_stp_truth_table}e. In this operation, currents from multiple rows accumulate on the \textit{RBLs} and are then subtracted to generate the MAC output. For clarity, Fig. \ref{fig_stp_truth_table}e illustrates the case of a single asserted row; however, in practice, multiple rows are asserted concurrently to exploit parallelism (as discussed later).

\begin{itemize}
    \item When the weight is '0', both cells are initialized with their PE capacitors polarized to \textit{-P}. Consequently, regardless of the input, both \textit{RBL1} and \textit{RBL2} conduct identical currents. Subtracting these yields zero current difference, corresponding to an output of '0'.
    \item When the input is '0', the lines applying the inputs to both cells are grounded. As a result, no current flows through either RBL, and the difference between the currents on \textit{RBL1} and \textit{RBL2} is zero. This also maps to an output of '0'.
    \item If the input is '1' and the weight is '1', the PeFET polarizations are (\textit{+P}, \textit{-P}), and a gate-to-body voltage $V_{GB} = V_{DD}/2$ is applied. This configuration produces $I_{LRS}$ in the \textit{+P} cell (Cell 1) and $I_{HRS}$ in the \textit{-P} cell (Cell 2). Subtracting these results in a current difference of $I_{LRS}-I_{HRS}$, which corresponds to an output of '1'.
    \item For input '1' and weight '-1', the polarizations are (\textit{-P}, \textit{+P}). The same $V_{GB} = V_{DD}/2$ now produces $I_{HRS}$ in Cell 1 and $I_{LRS}$ in Cell 2, resulting in a current difference of $I_{HRS}-I_{LRS}$, which corresponds to an output of '-1'.
    \item When the input is '-1', a negative $V_{GB} = -V_{DD}/2$ is applied to both cells. This exploits the unique $V_{GB}$-polarity and polarization-dependent current characteristics of PeFETs, causing the \textit{+P} cell to conduct $I_{HRS}$ and the \textit{-P} cell to conduct $I_{LRS}$. For a weight of '1' (\textit{+P}, \textit{-P}), the resulting current difference is $I_{HRS}-I_{LRS}$, corresponding to an output of '-1'. 
    \item For input '-1' and weight '-1' (\textit{-P}, \textit{+P}), the current difference becomes  $I_{LRS}-I_{HRS}$, corresponding to an output of '1'.
\end{itemize}

To enable subtraction of \textit{RBL} currents, we employ a comparator and an analog current subtractor as in \cite{niharika_fnano}, with the resultant current digitized by an ADC. Since ADC energy and latency often dominate, we assert 16 rows concurrently, following \cite{niharika_fnano}, to keep ADC overheads low. The comparator–subtractor–ADC chain was verified to operate robustly under 16-row parallel assertion \cite{niharika_fnano}.

\begin{figure}[!t]
\centering
\includegraphics[width=\columnwidth, height=2.05in]{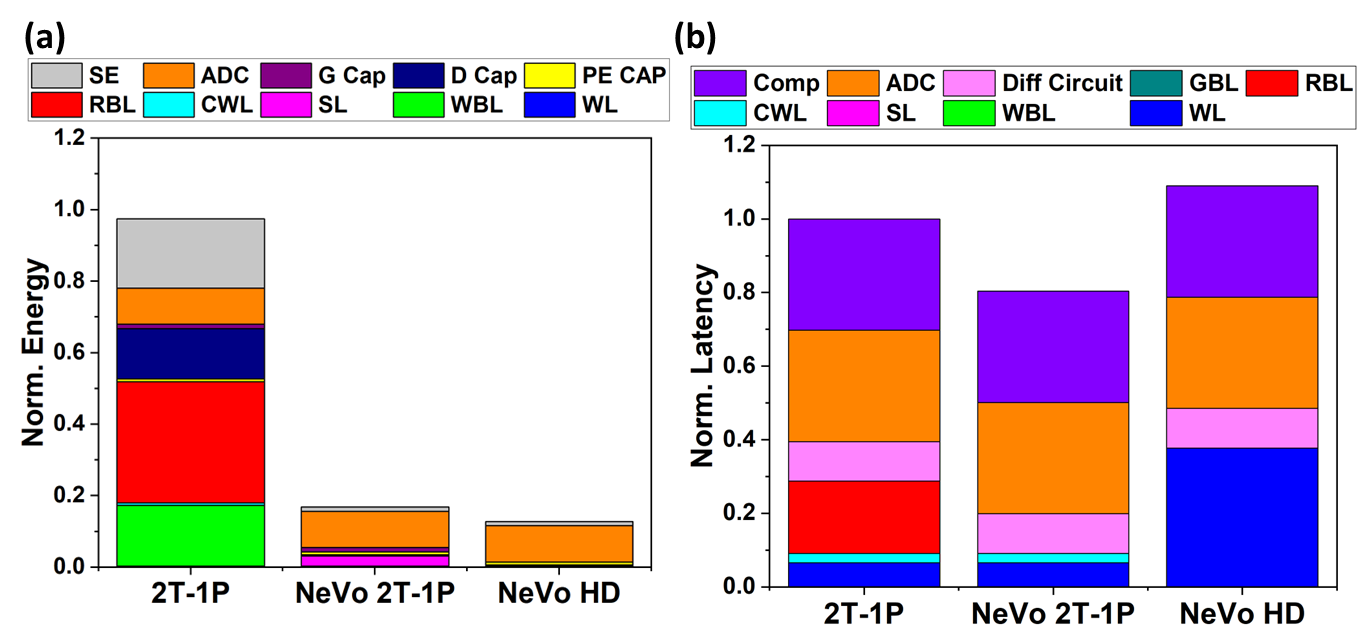}
\caption{Characterizing STP MAC using PeFET NVMs (a) Energy and (b) Latency comparison}
\label{fig_I_stp}
\end{figure}

Fig.\ref{fig_I_stp}a shows the energy and latency comparison for computing STP-MAC. NeVo 2T-1P IMC consumes only 0.19× the energy of 2T-1P IMC, primarily due to avoiding RBL and WBL assertions. Additionally, NeVo HD IMC consumes only 0.79× the energy of NeVo 2T-1P IMC and 0.15× that of 2T-1P IMC. This reduction stems from its more compact layout, the absence of gate capacitance overhead (from access transistors present in both NeVo 2T-1P and 2T-1P), and the avoidance of charging vertically running lines, as required in 2T-1P.

Fig.\ref{fig_I_stp}b shows that NeVo 2T-1P IMC achieves 0.8× the latency of 2T-1P IMC. This is primarily due to avoiding charging of \textit{RBL}, \textit{WBL} and their associated drain capacitances. Instead, NeVo 2T-1P asserts WL and SL which span only the segment width (as discussed in Section IV A). NeVo HD exhibits 1.36× (1.08x) the latency of NeVo 2T-1P (2T-1P) IMC. This is because, in NeVo HD, the \textit{WL} must charge the PE capacitance fully to $±V_{DD}/2$, whereas in NeVo 2T-1P, the access transistor between \textit{WBL} and the PeFET gate limits the \textit{CWL} charging to a lower magnitude. 

The work in \cite{niharika_fnano} has compared 2T-1P IMC with SRAM NMC and found that 2T-1P IMC saves 15\% energy and 91\% in terms of latency with respect to SRAM. Since NeVo HD IMC and NeVo 2T-1P IMC consume less energy and exhibit lower or comparable latency to 2T-1P IMC, we deduce that the NeVo approaches proposed in this paper (NeVo HD IMC and NeVo 2T-1P IMC) perform better than SRAM NMC. 

\section{Conclusion}
In this work, we introduced two PeFET NVM designs—NeVo HD and NeVo 2T-1P—that leverage negative-voltage biasing to eliminate costly bit-line charging. With optimized bit-cell and array layouts, these designs deliver substantial energy savings, particularly in read operations. Additionally, by exploiting the unique $V_{GB}$-polarity and polarization-controlled current of PeFETs (using the NeVo biasing approach), we implemented in-memory addition, subtraction, and STP-MAC primitives, achieving significant energy reductions over prior PeFET NVMs and SRAM.

\bibliographystyle{IEEEtran}
\bibliography{references}

@ARTICLE{niharika_ted,
  author={Thakuria, Niharika and Elangovan, Reena and Raghunathan, Anand and Gupta, Sumeet K.},
  journal={IEEE Transactions on Electron Devices}, 
  title={Piezoelectric Strain FET (PeFET)-Based Nonvolatile Memories}, 
  year={2023},
  volume={70},
  number={6},
  pages={3076-3084},
  doi={10.1109/TED.2023.3270845}}

@ARTICLE{niharika_fnano,
AUTHOR={Thakuria, Niharika  and Elangovan, Reena  and Thirumala, Sandeep K.  and Raghunathan, Anand  and Gupta, Sumeet K. },
TITLE={STeP-CiM: Strain-Enabled Ternary Precision Computation-In-Memory Based on Non-Volatile 2D Piezoelectric Transistors},
JOURNAL={Frontiers in Nanotechnology},
VOLUME={4},
YEAR={2022},
URL={https://www.frontiersin.org/journals/nanotechnology/articles/10.3389/fnano.2022.905407},
DOI={10.3389/fnano.2022.905407},
ISSN={2673-3013}}

@article{memory_wall,
author = {Wulf, Wm. A. and McKee, Sally A.},
title = {Hitting the memory wall: implications of the obvious},
year = {1995},
issue_date = {March 1995},
publisher = {Association for Computing Machinery},
address = {New York, NY, USA},
volume = {23},
number = {1},
issn = {0163-5964},
url = {https://doi.org/10.1145/216585.216588},
doi = {10.1145/216585.216588},
journal = {SIGARCH Comput. Archit. News},
month = mar,
pages = {20–24},
numpages = {5}
}

@ARTICLE{shubham_jain,
  author={Jain, Shubham and Ranjan, Ashish and Roy, Kaushik and Raghunathan, Anand},
  journal={IEEE Transactions on Very Large Scale Integration (VLSI) Systems}, 
  title={Computing in Memory With Spin-Transfer Torque Magnetic RAM}, 
  year={2018},
  volume={26},
  number={3},
  pages={470-483},
  doi={10.1109/TVLSI.2017.2776954}}

@INPROCEEDINGS{aziz_fefets,
  author={Aziz, Ahmedullah and Breyer, Evelyn T. and Chen, An and Chen, Xiaoming and Datta, Suman and Gupta, Sumeet Kumar and Hoffmann, Michael and Hu, Xiaobo Sharon and Ionescu, Adrian and Jerry, Matthew and Mikolajick, Thomas and Mulaosmanovic, Halid and Ni, Kai and Niemier, Michael and O'Connor, Ian and Saha, Atanu and Slesazeck, Stefan and Thirumala, Sandeep Krishna and Yin, Xunzhao},
  booktitle={2018 Design, Automation \& Test in Europe Conference \& Exhibition (DATE)}, 
  title={Computing with ferroelectric FETs: Devices, models, systems, and applications}, 
  year={2018},
  volume={},
  number={},
  pages={1289-1298},
  doi={10.23919/DATE.2018.8342213}}

@ARTICLE{magic,
  author={Kvatinsky, Shahar and Belousov, Dmitry and Liman, Slavik and Satat, Guy and Wald, Nimrod and Friedman, Eby G. and Kolodny, Avinoam and Weiser, Uri C.},
  journal={IEEE Transactions on Circuits and Systems II: Express Briefs}, 
  title={MAGIC—Memristor-Aided Logic}, 
  year={2014},
  volume={61},
  number={11},
  pages={895-899},
  doi={10.1109/TCSII.2014.2357292}}

@INPROCEEDINGS{jeffry_icecs,
  author={Louis, Jeffry and Hoffer, Barak and Kvatinsky, Shahar},
  booktitle={2019 26th IEEE International Conference on Electronics, Circuits and Systems (ICECS)}, 
  title={Performing Memristor-Aided Logic (MAGIC) using STT-MRAM}, 
  year={2019},
  volume={},
  number={},
  pages={787-790},
  doi={10.1109/ICECS46596.2019.8965179}}

@ARTICLE{adra,
  author={Malhotra, Akul and Saha, Atanu K. and Wang, Chunguang and Gupta, Sumeet K.},
  journal={IEEE Transactions on Circuits and Systems II: Express Briefs}, 
  title={ADRA: Extending Digital Computing-In-Memory With Asymmetric Dual-Row-Activation}, 
  year={2023},
  volume={70},
  number={8},
  pages={3089-3093},
  doi={10.1109/TCSII.2023.3253659}}

@ARTICLE{sram_negative_bl,
  author={Mukhopadhyay, Saibal and Rao, Rahul M. and Kim, Jae-Joon and Chuang, Ching-Te},
  journal={IEEE Transactions on Very Large Scale Integration (VLSI) Systems}, 
  title={SRAM Write-Ability Improvement With Transient Negative Bit-Line Voltage}, 
  year={2011},
  volume={19},
  number={1},
  pages={24-32},
  doi={10.1109/TVLSI.2009.2029114}}

@INPROCEEDINGS{sram_write_assist,
  author={Chang, Jonathan and Chen, Yen-Huei and Cheng, Hank and Chan, Wei-Min and Liao, Hung-Jen and Li, Quincy and Chang, Stanley and Natarajan, Sreedhar and Lee, Robin and Wang, Ping-Wei and Lin, Shyue-Shyh and Wu, Chung-Cheng and Cheng, Kuan-Lun and Cao, Min and Chang, George H.},
  booktitle={2013 IEEE International Solid-State Circuits Conference Digest of Technical Papers}, 
  title={A 20nm 112Mb SRAM in High-k metal-gate with assist circuitry for low-leakage and low-VMIN applications}, 
  year={2013},
  volume={},
  number={},
  pages={316-317},
  doi={10.1109/ISSCC.2013.6487750}}

@article{PTM,
author = {Zhao, Wei and Cao, Yu},
title = {Predictive technology model for nano-CMOS design exploration},
year = {2007},
issue_date = {April 2007},
publisher = {Association for Computing Machinery},
address = {New York, NY, USA},
volume = {3},
number = {1},
issn = {1550-4832},
url = {https://doi.org/10.1145/1229175.1229176},
doi = {10.1145/1229175.1229176},
journal = {J. Emerg. Technol. Comput. Syst.},
month = apr,
pages = {1–es},
numpages = {17}
}

@inproceedings{technology_scaling,
author = {Shalf, John and Dosanjh, Sudip and Morrison, John},
title = {Exascale computing technology challenges},
year = {2010},
isbn = {9783642193279},
publisher = {Springer-Verlag},
address = {Berlin, Heidelberg},
booktitle = {Proceedings of the 9th International Conference on High Performance Computing for Computational Science},
pages = {1–25},
numpages = {25},
location = {Berkeley, CA},
series = {VECPAR'10}
}

@inproceedings{Ambit,
author = {Seshadri, Vivek and Lee, Donghyuk and Mullins, Thomas and Hassan, Hasan and Boroumand, Amirali and Kim, Jeremie and Kozuch, Michael A. and Mutlu, Onur and Gibbons, Phillip B. and Mowry, Todd C.},
title = {Ambit: in-memory accelerator for bulk bitwise operations using commodity DRAM technology},
year = {2017},
isbn = {9781450349529},
publisher = {Association for Computing Machinery},
address = {New York, NY, USA},
url = {https://doi.org/10.1145/3123939.3124544},
doi = {10.1145/3123939.3124544},
booktitle = {Proceedings of the 50th Annual IEEE/ACM International Symposium on Microarchitecture},
pages = {273–287},
numpages = {15},
location = {Cambridge, Massachusetts},
series = {MICRO-50 '17}
}

@INPROCEEDINGS{geniex,
  author={Chakraborty, Indranil and Fayez Ali, Mustafa and Eun Kim, Dong and Ankit, Aayush and Roy, Kaushik},
  booktitle={2020 57th ACM/IEEE Design Automation Conference (DAC)}, 
  title={GENIEx: A Generalized Approach to Emulating Non-Ideality in Memristive Xbars using Neural Networks}, 
  year={2020},
  volume={},
  number={},
  pages={1-6},
  doi={10.1109/DAC18072.2020.9218688}}

@ARTICLE{x_sram,
  author={Agrawal, Amogh and Jaiswal, Akhilesh and Lee, Chankyu and Roy, Kaushik},
  journal={IEEE Transactions on Circuits and Systems I: Regular Papers}, 
  title={X-SRAM: Enabling In-Memory Boolean Computations in CMOS Static Random Access Memories}, 
  year={2018},
  volume={65},
  number={12},
  pages={4219-4232},
  doi={10.1109/TCSI.2018.2848999}}

@INPROCEEDINGS{sandeep_ternary,
  author={Thirumala, Sandeep Krishna and Jain, Shubham and Gupta, Sumeet Kumar and Raghunathan, Anand},
  booktitle={2020 Design, Automation \& Test in Europe Conference \& Exhibition (DATE)}, 
  title={Ternary Compute-Enabled Memory using Ferroelectric Transistors for Accelerating Deep Neural Networks}, 
  year={2020},
  volume={},
  number={},
  pages={31-36},
  doi={10.23919/DATE48585.2020.9116495}}

@ARTICLE{tim_dnn,
  author={Jain, Shubham and Gupta, Sumeet Kumar and Raghunathan, Anand},
  journal={IEEE Transactions on Very Large Scale Integration (VLSI) Systems}, 
  title={TiM-DNN: Ternary In-Memory Accelerator for Deep Neural Networks}, 
  year={2020},
  volume={28},
  number={7},
  pages={1567-1577},
  doi={10.1109/TVLSI.2020.2993045}}

@inproceedings{precision1,
author = {Courbariaux, Matthieu and Bengio, Yoshua and David, Jean-Pierre},
title = {BinaryConnect: training deep neural networks with binary weights during propagations},
year = {2015},
publisher = {MIT Press},
address = {Cambridge, MA, USA},
booktitle = {Proceedings of the 29th International Conference on Neural Information Processing Systems - Volume 2},
pages = {3123–3131},
numpages = {9},
location = {Montreal, Canada},
series = {NIPS'15}
}

@inproceedings{precision2,
author = {Colangelo, Philip and Nasiri, Nasibeh and Nurvitadhi, Eriko and Mishra, Asit and Margala, Martin and Nealis, Kevin},
title = {Exploration of Low Numeric Precision Deep Learning Inference Using Intel® FPGAs: (Abstract Only)},
year = {2018},
isbn = {9781450356145},
publisher = {Association for Computing Machinery},
address = {New York, NY, USA},
url = {https://doi.org/10.1145/3174243.3174999},
doi = {10.1145/3174243.3174999},
pages = {294},
numpages = {1},
location = {Monterey, CALIFORNIA, USA},
series = {FPGA '18}
}

@inproceedings{
tnn,
title={{WRPN}: Wide Reduced-Precision Networks},
author={Asit Mishra and Eriko Nurvitadhi and Jeffrey J Cook and Debbie Marr},
booktitle={International Conference on Learning Representations},
year={2018},
url={https://openreview.net/forum?id=B1ZvaaeAZ},
}

@INPROCEEDINGS{edram,
  author={Yoo, Taegeun and Kim, Hyunjoon and Chen, Qian and Kim, Tony Tae-Hyoung and Kim, Bongjin},
  booktitle={2019 IEEE/ACM International Symposium on Low Power Electronics and Design (ISLPED)}, 
  title={A Logic Compatible 4T Dual Embedded DRAM Array for In-Memory Computation of Deep Neural Networks}, 
  year={2019},
  volume={},
  number={},
  pages={1-6},
  doi={10.1109/ISLPED.2019.8824826}}

@INPROCEEDINGS{presiach,
  author={Saha, Atanu K. and Gupta, Sumeet K.},
  booktitle={2018 76th Device Research Conference (DRC)}, 
  title={Modeling and Comparative Analysis of Hysteretic Ferroelectric and Anti-ferroelectric FETs}, 
  year={2018},
  volume={},
  number={},
  pages={1-2},
  doi={10.1109/DRC.2018.8442136}}

@article{experimental,
    author = {Malakooti, Mohammad H. and Sodano, Henry A.},
    title = {Noncontact and simultaneous measurement of the d33 and d31 piezoelectric strain coefficients},
    journal = {Applied Physics Letters},
    volume = {102},
    number = {6},
    pages = {061901},
    year = {2013},
    month = {02},
    issn = {0003-6951},
    doi = {10.1063/1.4791573},
    url = {https://doi.org/10.1063/1.4791573},
    eprint = {https://pubs.aip.org/aip/apl/article-pdf/doi/10.1063/1.4791573/14277895/061901\_1\_online.pdf},
}

@article{s2ds,
    author = {Suryavanshi, Saurabh V. and Pop, Eric},
    title = {S2DS: Physics-based compact model for circuit simulation of two-dimensional semiconductor devices including non-idealities},
    journal = {Journal of Applied Physics},
    volume = {120},
    number = {22},
    pages = {224503},
    year = {2016},
    month = {12},
    issn = {0021-8979},
    doi = {10.1063/1.4971404},
    url = {https://doi.org/10.1063/1.4971404},
    eprint = {https://pubs.aip.org/aip/jap/article-pdf/doi/10.1063/1.4971404/15185757/224503\_1\_online.pdf},
}

@ARTICLE{supreet_sram,
  author={Jeloka, Supreet and Akesh, Naveen Bharathwaj and Sylvester, Dennis and Blaauw, David},
  journal={IEEE Journal of Solid-State Circuits}, 
  title={A 28 nm Configurable Memory (TCAM/BCAM/SRAM) Using Push-Rule 6T Bit Cell Enabling Logic-in-Memory}, 
  year={2016},
  volume={51},
  number={4},
  pages={1009-1021},
  doi={10.1109/JSSC.2016.2515510}}

@article{feram_seg,
author = {J√ºrgen T. Rickes and Hugh Mcadams and James Grace and John Fong and Steve Gilbert and Angela Wang and Dave Lee and Cezary Pietrzyk and Ralph Lanham and Jun Amano and Scott Summerfelt and Ted Moise and Rainer M. Waser and},
title = {A Novel Sense-Amplifier and Plate-Line Architecture for Ferroelectric Memories},
journal = {Integrated Ferroelectrics},
volume = {48},
number = {1},
pages = {109--118},
year = {2002},
publisher = {Taylor \& Francis},
doi = {10.1080/713718311}
}

\end{document}